\documentclass[%
 reprint,
 amsmath,amssymb,
 aps,
]{revtex4-1}

\usepackage{graphicx}
\usepackage[space]{grffile}
\usepackage{latexsym}
\usepackage{textcomp}
\usepackage{longtable}
\usepackage{multirow,booktabs}
\usepackage{amsfonts,amsmath,amssymb}
\usepackage{natbib}
\usepackage{url}
\usepackage{hyperref}
\hypersetup{colorlinks=false,pdfborder={0 0 0}}
\usepackage[utf8]{inputenc}
\usepackage[english]{babel}
\begin{document}

\title{Electronic Band Structure Effects in the Stopping of Protons in Copper}

\author{Edwin E. Quashie}
\affiliation{Department of Physics, Florida A\&M University, Tallahassee, FL 32307, USA}
 
\author{Bidhan C. Saha}
\affiliation{Department of Physics, Florida A\&M University, Tallahassee, FL 32307, USA}
 
\author{Alfredo A. Correa}
\affiliation{Lawrence Livermore National Laboratory, 7000 East Avenue, Livermore, CA 94550, USA}

\begin{abstract}
We present an ab initio study of the electronic stopping power of protons in copper over a wide range of proton velocities $v = 0.02-10~\mathrm{a.u.}$ where we take into account non-linear effects.
Time-dependent density functional theory coupled with molecular dynamics is used to study electronic excitations produced by energetic protons.
A plane-wave pseudopotential scheme is employed to solve the time-dependent Kohn-Sham equations for a moving ion in a periodic crystal.
The electronic excitations and the band structure determine the stopping power of the material and alter the interatomic forces for both channeling and off-channeling trajectories. 
Our off-channeling results are in quantitative agreement with experiments, and at low velocity they unveil a crossover region of superlinear velocity dependence (with a power of $\sim 1.5$) in the velocity range $v = 0.07-0.3~\mathrm{a.u.}$, which we associate to the copper crystalline electronic band structure.
The results are rationalized by simple band models connecting two separate regimes. 
We find that the limit of electronic stopping $v\to 0$ is not as simple as phenomenological models suggest and it plagued by band-structure effects.

\end{abstract}%

\maketitle

\section{Introduction}

The interaction of charged particles with matter has been a subject of extensive research over many decades. 
These studies provide information for many technological applications such as nuclear safety, 
applied material science, medical physics and fusion and fission applications 
\cite{Komarov_2013,Patel_2003,Caporaso_2009,Odette_2005}.

Among the measurable quantities associated to the interaction between ions and solids, the stopping power $\mathrm(S)$ \cite{Ferrell_1977} has received much attention; it provides information regarding the energy transfer between the incoming projectile and the solid target.
When a fast ion moves through a material, it loses most of its kinetic energy due to the excitations of the target electrons along its trajectory in what constitutes a fundamentally non-adiabatic process.
This energy-loss phenomenon plays an important role in many experimental studies involving radiation in solids, surfaces, and nanostructures \cite{Chenakin_2006,Figueroa_2007,Markin_2008,Kaminsky_1965,Lehmann_1978,Sigmund_2014,Nastasi_1996}. 

Various models and theories have been proposed to calculate stopping cross sections due to electrons. 
Employing the First Born Approximation, Bethe \cite{Bethe_1930_EN} has introduced the first calculations of inelastic and ionization cross section. The Bloch correction \cite{Bloch_1933} provides a convenient link between the Bohr and the Bethe scheme. 
Fermi and Teller \cite{Fermi_1947} using electron gas models had reported electronic stopping for various targets. 
The Bethe formula for stopping has been studied in details by Lindhard and Winther \cite{Lindhard_Winther} on the basis of the generalized linear-response theory. 
%
%In the low energy region the energy loss in metal is due to the excitation of a portion of electrons around the Fermi level to empty states in the conducting band at higher energies, a minimum momentum transfer due to the limited response time of target electrons to the projectile ions. In this paper we concentrate in the near and below the maximum of stopping down to the low velocity regime for $\mathrm{H}$ in $\mathrm{Cu}$.
%

In particular, the condensed matter community has introduced sophisticated numerical computer simulation techniques for this fundamentally non-adiabatic problem as spearheaded by Echenique \emph{et al.} ~\cite{Echenique_1989} aimed to overcome limitations of historical approaches \cite{Bethe_1930,Bloch_1933,Fermi_1947,Lindhard_Winther}.
A unified \emph{ab initio} theoretical approach suitable for different projectiles and energies is in its developing stages \cite{Pruneda_2007,Schleife_2015,Ullah_2015}. 
A review on the topic can be found in Ref.~\cite{Race_2010} and references therein.

Using a Kohn-Sham (KS) scheme of time-dependent density functional theory (TDDFT) where the KS wave functions are expanded in a basis set of spherical harmonics, Quijada \emph{et al.} \cite{Quijada_2007} have studied the energy loss of protons and anti-protons moving inside metallic spherical $\mathrm{Al}$ (Jellium) clusters and obtained good results for the projectile-target energy transfer over a restricted energy range. 
Recently Uddin \emph{et al.} \cite{Alfaz_Uddin_2013} and Haque \emph{et al.} \cite{Haque_2015} have calculated stopping cross sections for various media using atomic density functions from Dirac-Hartree-Fock-Slater wave functions in the Lindhard-Schraff theory \cite{Lindhard_Scharff} with fitted parameters and obtained close agreement with the experimental and \textsc{Srim} data.
\textsc{Srim} \cite{Ziegler_2010} provides both a fitted model for electronic stopping as well as a large set of experimental points.

In studying the role of ion-solid interactions of $\mathrm{H^+}$ in $\mathrm{Al}$, Correa \emph{et al.} \cite{Correa_2012} have shown that the electronic excitations affect the interatomic forces relative to the adiabatic outcome. 
Recently, Schleife \emph{et al.} \cite{Schleife_2015} have calculated the electronic stopping power ($S_\text{e}$) of $\mathrm{H}$ and $\mathrm{He}$ projectiles including 
TDDFT non-adiabatic electron dynamics and found that off-channeling trajectories \emph{along with} the inclusion of semicore electrons enhance $S_\text{e}$, resulting in much better agreement with the \textsc{Srim} experimental and modeled data \cite{Ziegler_2010} in a wide range of energies. 
In this case we concentrate in a metal with a richer electronic band structure around the Fermi energy, such as $\mathrm{Cu}$.

The recent measurements by Cantero \emph{et al.} \cite{Cantero_2009} and by Markin \emph{et al.} \cite{Markin_2009} of slow ($v \leq 0.6~\mathrm{a.u.}$) $\mathrm{H^+}$ in $\mathrm{Cu}$ give a glimpse of the interesting extreme low velocity limit. 
Although disagreeing with each other in absolute scale by $\sim 40\%$ (Fig. \ref{fig:stopping_power}), both reveal the stopping due to conduction band electronic excitations at lower velocity, evidenced as a change in slope near $v=0.15$ or $0.10~\mathrm{a.u.}$ respectively. 
The change of slope was deduced qualitatively to be caused by the participation of $\mathrm{d}$-electrons \cite{Goebl_2013}.

In this paper we will address the problem of theoretical calculation of $S_\text{e}$ of protons in crystalline $\mathrm{Cu}$ for a wide range of available experimental velocities ($0.02~\mathrm{a.u.} \leq v \leq 10~\mathrm{a.u.}$). 
We perform our calculations by directly simulating the process of a proton traversing a crystal of $\mathrm{Cu}$ atoms, producing individual and collective electronic excitations within the TDDFT framework \cite{Correa_2012,Schleife_2012,Schleife_2014} including Ehrenfest molecular dynamics (EMD) \cite{Gross_1996,Calvayrac_2000,Mason_2007,Alonso_2008,Andrade_2009}. This method is used to calculate most microscopic quantities along the process (forces, electronic density, charges, etc); in particular, we concentrate here in the calculation of $S_\text{e}$. 
A quantitative explanation and interpretation of our results are furnished along with a detailed experimental comparison.

\section{Method}

%.

During the course of the simulation, we monitored the energy transferred to the electrons of the target due to a constant velocity moving proton. 
%At the time scales of the simulations, the large mass of the proton guarantees a change in its velocity that is relatively small.
For simplicity, and since the eletronic stopping is a velocity-resolved quantity the proton is constrained to move at constant velocity, hence the total energy of the system is not conserved. 
The excess in total energy is instead used as a measure of the stopping power as a function of the proton velocity.
As the proton moves, the time-dependent Kohn-Sham (TDKS) equation \cite{Runge_1984} describes the evolution of the electronic density and energy of the system, due to the dynamics of effective single-particle states under the external potential generated by the proton and the crystal of $\mathrm{Cu}$ nuclei. 
These states are evolved in time with a self-consistent Hamiltonian that is a functional of the density:
\begin{equation}
    \mathrm i\hbar\tfrac\partial{\partial t}\psi_i(\mathbf{r}, t) = \left\{-\tfrac{\hbar^2\nabla^2}{2m} + V_\text{KS}[n(t), \{\mathbf{R}_J(t)\}_J](\mathbf{r}, t)\right\}\psi_i(\mathbf{r}, t)
\label{eq:tdks1} 
\end{equation}

The KS effective potential $V_\text{KS}[n(t), \{\mathbf {R}_J(t)\}_J](\mathbf{r}, t)$ is given by
\begin{equation}
\begin{aligned}
V_\text{KS}[n, \{\mathbf{R}_J(t)\}_J] = V_\text{ext}[\{\mathbf{R}_J(t)\}_J] + V_\text{H}[n] +  V_\text{XC}[n] 
\label{eq:tdks3}
\end{aligned}
\end{equation}

where the external potential is $V_\text{ext}[\{\mathbf{R}_J(t)\}_J](\mathbf{r}, t)$ due to ionic core potential (with ions at positions $\mathbf R_J(t)$), $V_\text{H}[n](\mathbf{r}, t)$ is the Hartree potential comprising the classical electrostatic interactions between electrons and $V_\text{XC}[n](\mathbf{r}, t)$ denotes the exchange-correlation (XC) potential. 
The spatial and time coordinates are represented by $\mathbf{r}$ and $t$ respectively. 
At time $t$ the instantaneous density is given by the sum of individual electron probabilities $n(\mathbf{r}, t) = \sum_i |\psi_i(\mathbf{r}, t)|^2$.
The XC potential used in this study is due to Perdew-Burke-Ernzerhof (PBE) ~\cite{Perdew_1992,Perdew_1996}, and we used norm-conserving Troullier-Martins pseudopotential to represent $V_\text{ext}$, with $17$ explicit electrons per $\mathrm{Cu}$ atom (not necessarily all 17 electrons participate in the process as we will discuss later).

Each simulation of the ion-solid collision consists of a well-defined trajectory of the projectile in the FCC metallic bulk sample with experimental density. 
The calculations were done using the code \textsc{Qbox} \cite{Gygi_2008} with time-dependent modifications \cite{Schleife_2012}\cite{Draeger_2016}. 
The KS orbitals are expanded in a supercell plane-wave basis. 
The advantages of using the plane-wave approach is that it systematically deals with basis-size effects, which was a drawback for earlier approaches \cite{Pruneda_2007, Correa_2012}.

Periodic boundary conditions along with Ewald summation~\cite{Amisaki_2000,Roy_2007} are used throughout this study. 
The supercell size was chosen so as to reduce the specious size effects while maintaining controllable computational demands. 
Since the larger size effects are negligible this calculation used $(3\times3\times3)$ conventional cells containing $108$ host $\mathrm{Cu}$ atoms and $\mathrm{H^+}$.
% also represented by a Troullier-Martins pseudopotential ($17$ valence electrons per copper atom are explicitly considered). 
To integrate the Brillouin zone a single $k$-point ($\Gamma$) was used, except for test cases.
The screening length of $\mathrm{Cu}$ is close to the interatomic spacing, which reduces the range of Coulomb interactions and makes it controllable in a periodic representation.

Finite size effects are studied between 108 and 256 atoms in a supercell of $(3\times3\times3)$ and $(4\times4\times4)$ respectively in this simulation, the errors remain within 3\% in conformity with the earlier observation \cite{Schleife_2015}.

The plane-wave basis set is sampled accurately with a $130~\mathrm{Ry}$ energy cutoff.  
We also tested for k-point convergence in a $(3\times3\times3)$-Monkhorst-Pack grid (for the cubic 108-supercell), which would be equivalent to a 2916 
($108\times 27$ simulation cell of an hypothetical periodic system, including replicas of the proton), for selected velocities with negligible differences within 0.08\%.

The projectile $\mathrm{H^+}$ is initially placed in the crystal and a time-\emph{independent} DFT calculation was completed to obtain the converged ground state results that are required as the initial condition for subsequent evolution with the moving projectile.
We then perform TDDFT calculations on the electronic system with the moving proton in the channeling and off-channeling geometries.
Following the method introduced by Pruneda \emph{et al.} \cite{Pruneda_2005} the projectile is put in motion with a constant velocity in a straight along a $\langle 100\rangle$ channeling trajectory (also called hyper-channeling) which minimizes the collision of the projectile with the host atoms~\cite{Pruneda_2007,Correa_2012,Schleife_2015}. 

In the off-channeling case the projectile takes random trajectory directions through the host crystal yielding occasionally stronger interaction between the projectile and the tightly bound electrons of the host atom. 
The use of off-channeling trajectories was introduced in Ref.~\cite{Schleife_2015}.
%Finite-size errors in the simulations are overcome by considering large simulation cells \cite{Schleife_2015}.%

The fourth-order Runge-Kutta scheme (RK4) \cite{Schleife_2012} is used to propagate the electronic orbitals in time, 
following the scheme of Ref.~\cite{Schleife_2012} the TDKS equation (see Eq. \ref{eq:tdks1})  
with a time step of, at most, $0.121~\mathrm{attoseconds}$ 
(which is within the numerical stability limit time-integration scheme for this basis set). 
High velocity points were simulated with smaller time steps.
The wavefunctions were then propagated for up to tens of femtoseconds.

The total electronic energy ($E$) of the system changes as a function of the projectile position ($x$) since the projectile 
(forced to maintain its velocity) deposits energy into the electronic system as it moves through the host atoms. 
The increase of $E$ as a function of projectile displacement $x$ enables us to extract the electronic stopping power as a time-averaged quantity for each velocity,
\begin{equation}
S_\text{e} = \overline{\mathrm{d}E(x)/\mathrm{d}x}
\label{eq:stopping}  
\end{equation}
$S_\text{e}$ has the dimension of a force (e.g. $E_\text{h}/a_0$) and it has the interpretation of a drag force acting on the projectile.

\section{Results}

\begin{figure}[h!]
\begin{center}
\includegraphics[width=1\columnwidth]{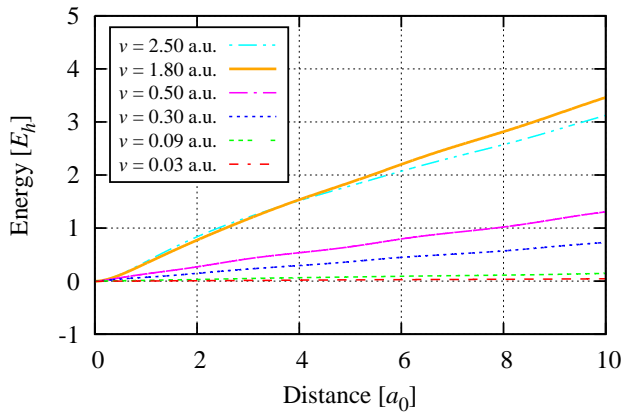}
\caption{{\label{fig:energy_distance} (color online). Total energy increase as a function of position for a proton in crystalline $\mathrm{Cu}$.
The lattice atoms are kept at their equilibrium positions while the projectile passes in a $\langle 100\rangle$ channeling trajectory at velocity $v$.%%
}}
\end{center}
\end{figure}

%\section{Results}

Our calculations of $\mathrm{H^+}\to \mathrm{Cu}$ system in the range of velocities between $0.02$ and $10~\mathrm{a.u.}$ 
are presented in details in this section.
Fig. \ref{fig:energy_distance} shows the total electronic energy of the $\mathrm{H^+ + Cu}$ system as a function of position for various projectile velocities for the hyper-channeling case. 
At a lower velocities regime, the energy transfer is smaller, approaching the adiabatic behavior, while at higher velocities (aside oscillations of the total energy with the position of the projectile) the total energy of the system increases linearly with time.
After the projectile travels some short distance in the crystals ($\sim 3~a_0$) the increase in total energy of the system stabilizes to a steady rate. 
At that steady state, the $S_\text{e}$ is then extracted from the average slope of the total energy vs. projectile displacement; 
it represents the rate of energy gained by the target and loss by the projectile.

\begin{figure}[h!]
\begin{center}
\includegraphics[width=1\columnwidth]{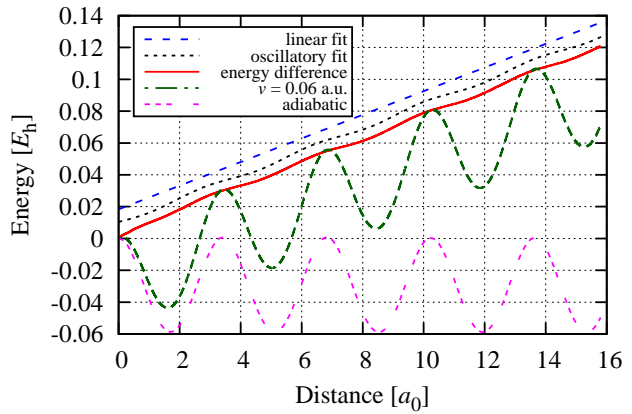}
\caption{{\label{fig:fit_graph} (color online). The increase of $E$ as a function of proton position at $v = 0.06 ~\mathrm{a.u.}$ (green dashed line) along $\langle 100\rangle$ channeling trajectory and the adiabatic energy (magenta dot-dashed line). 
The adiabatic curve would correspond to a proton moving infinitely slowly, where there is no transfer of energy, just oscillations of the total energy with an overall zero slope.
%The initial (ground state energy) is subtracted. 
The oscillations in the curves reflect the periodicity of the $\mathrm{Cu}$ lattice. 
The red solid line shows the energy difference (subtraction of adiabatic energy from the $v = 0.06 ~\mathrm{a.u.}$ curves). 
The blue dashed and black dotted lines shows the slope of linear and oscillatory fits of the red solid line 
from $x = 5.0~a_0$ (after the transient) to a given maximum position $x$ as a function of this maximum position respectively. 
For visualization purposes the black line has been shifted vertically. %
%$v = 0 ~\mathrm{a.u.}$ from $v = 0.06 ~\mathrm{a.u.}$ curves)%
%%should be added a constant value of $-16856.62389 ~\mathrm{a.u.}$.%
A linear fit , $y = a + bx$ (blue line) yields a slope of $6.989 \times 10^{-3}~E_\text{h}/a_0$ with an error of $\pm 6.936 \times 10^{-5}~E_\text{h}/a_0$. 
We then proceed for a linear fit in addition to an oscillatory function $y = a + bx + A\cos(k x + \phi)$ (black dotted line) to capture any remnant oscillation. 
This oscillatory fit generates a slope of $7.435 \times 10^{-3}~E_\text{h}/a_0$ with an error of $\pm 8.52 \times 10^{-7}~E_\text{h}/a_0$, that is a minimal fitting error is obtained in the \emph{channeling} trajectory.%
}}
\end{center}
\end{figure}

In the low velocity cases extracting a slope becomes more challenging, 
first because a longer time simulation is required to sample the crystalline structure and second because the natural oscillations associated with the crystal periodicity becomes relatively larger. 
For the channeling cases, we subtract the adiabatic energy from time-dependent energy and we perform an oscillatory fit to the resulting curve. 
Fig. \ref{fig:fit_graph} explains how the slopes of Fig. \ref{fig:energy_distance} are evaluated in these cases. 
For higher velocities, a linear fit alone is enough to obtain reasonable values with small relative errors.%

\begin{figure}[h!]
\begin{center}
\includegraphics[width=1\columnwidth]{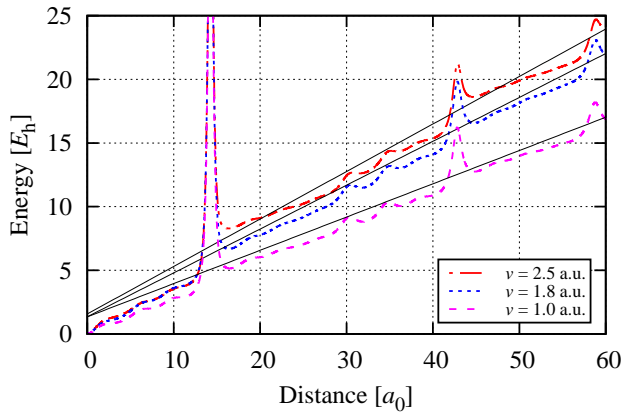}
\caption{{\label{fig:fit_off_channel} (color online) The total energy increase as a function of proton position at velocities $v = 1.0 ~\mathrm{a.u.}$ (magenta line), $v = 1.8 ~\mathrm{a.u.}$ (blue line) and $v = 2.5 ~\mathrm{a.u.}$ (red line) along the off-channeling direction [$0.309, 0.5, 0.809$] trajectory with the corresponding linear fits (black solid lines).%
}}
\end{center}
\end{figure}

For the off-channeling trajectory, the procedure for computing the $S_\text{e}$ is depicted in Fig. \ref{fig:fit_off_channel}. 
We used two directions, approximately [$0.705, 0.610, 0.363$] and [$0.309, 0.5, 0.809$] 
(given normalized here). 
The first direction was chosen by visual inspection of the supercell in order to \emph{not} match any simple channel but also avoid an immediate head on collision.
The second direction is the normalized version of $[1, \phi, \phi^2]$, where $\phi$ is the golden ratio ($\sim 1.618$), which guarantees a trajectory most incommensurate with the cell due to its mathematical properties as an irrational number.
It is important to note here an interesting geometrical fact that, for a direction incommensurate with the crystal directions, all available densities and impact parameters (distances of closest approach to host atom) are probed (averaged) eventually for a long enough trajectory.
Our simulations are limited in space (and time) but it is clear that the trajectories explore a wide range of impact parameters and therefore densities.
The viability and the necessity of considering this geometrical averaging method was shown earlier in \cite{Schleife_2015}. 

In Fig.~\ref{fig:fit_off_channel}, the sharp peaks show when the proton is in the vicinity of a host $\mathrm{Cu}$ atom during an off-channeling trajectory, 
while the smaller peaks and flatter regions indicate that the proton is not very close to any host atom. 
To obtain the $S_\text{e}$ we compute the slopes of the curves by a linear fit of the form $y = a + bx$ (black solid lines) using our data from $x > 5~a_0$ (to eliminate the transient region) up to a given maximum position of $x$ determined by minimizing reentrancy in the periodic supercell near the initial position. 
The slope ($b$) gives the electronic stopping for this off-channeling case.
%We have displayed the procedure used in obtaining the electronic stopping for the off-channeling trajectory case in Figure \ref{fig:fit_off_channel}.%  

\begin{figure}[h!]
\begin{center}
\includegraphics[width=1\columnwidth]{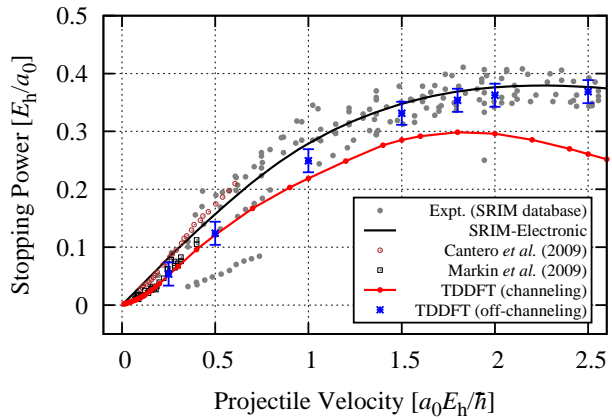}
\caption{{\label{fig:stopping_power} (color online). 
The $S_\text{e}$ vs. projectile velocity $v$: for a channeling trajectory (red lines) and for off-channeling trajectory (blue crosses). 
The black continuous line refers to the nominal tabulated model from \textsc{Srim}, based on its database for electronic stopping power \cite{Ziegler_2010}. 
The gray dots refer to the \textsc{Srim} experimental database contained therein \cite{Copper_Stop}. 
The error bars of the off-channeling trajectory are shown by the blue vertical lines, while error bars in channeling trajectories are smaller than the points.
Open circles and open squares show data of Cantero \emph{et al.} \cite{Cantero_2009} and Markin \emph{et al.} \cite{Markin_2009} respectively. 
}}
\end{center}
\end{figure}

Fig. \ref{fig:stopping_power} shows a comparison of our calculated $S_\text{e}$ with \textsc{Srim}-based model and experimental data. 
In the \emph{channeling} case, the maximum of our calculated stopping is lower in value and velocity compared to the \textsc{Srim} database and the \emph{off-channeling} case, and it decreases faster after the maximum.

For the off-channeling case, there is a better agreement between our $S_\text{e}$ results with the \textsc{Srim} data in most of the range. 
In experiments, where trajectories are not necessarily finely controlled, the projectile does indeed explore core regions of the host atoms, and that is presumably why off-channeling simulations are a better representation for the most common experiments \cite{Dorado_1993}. 
At higher velocities ($v > 4 ~\mathrm{a.u.}$) further disagreement stems from combined effect of the lack of explicit deeper core electrons in the simulation and also size effects, as excitations of long wavelength plasma oscillations are artificially constrained by the simulation supercell \cite{Schleife_2015}.
It is clear that a larger cell and eventually the inclusion of more core electrons would be necessary to obtain better agreement in this high velocity region that is out the of scope of this article.

Although experimental values have considerable vertical spread, 
our calculated stopping power is on the low side for most of the points and also below the fitted by \textsc{Srim} model \cite{Ziegler_2010}.
While this was partially explained by taking into account off-channeling trajectories near the maximum of stopping, 
there are other possible intrinsic limitations of the approximations to the density functional theory used.
Along this line, we would like to note that more sophisticated approaches, 
based on the dielectric and current-density response but including
the exact many-body and dynamic exchange-correlation
treatment, are available in the literature \cite{Nazarov_2007}. 
This type of advanced approaches which are beyond the current scope 
contain explicitly additional channels of dissipation not taken into account 
in our adiabatic exchange and correlation functionals, which can be relatively important.
Given the aforementioned limitations of the orbital based method and the exchange and correlation used it 
is still reassuring to see agreement up to a few times the velocity of the maximum stopping and gives us confidence to make predictions in the lower velocity regime.

At low velocity we observe that the off-channeling and channeling simulated results collapse into a common curve, 
this effect has been seen in the simulations before \cite{Correa_2012,Schleife_2015,Ullah_2015}. 
The effect is that the computed quantity less sensitive to the precise geometry of the trajectory,
as the geometric cross section increases. 
We speculate that this is because the effective binary cross section increases beyond the interatomic separation making the energy loss less sensitive the precise geometry of the environment.

\section{Discussion}

\begin{figure}[h!]
\begin{center}
\includegraphics[width=1\columnwidth]{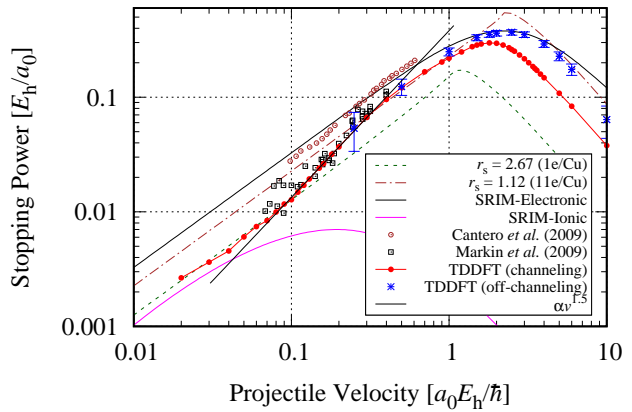}
\caption{{\label{fig:log_stopping_power} (color online)
The average $S_\text{e}$ vs. projectile velocity $v$: for a channeling trajectory (red circles) 
and for off-channeling trajectory (blue crosses). 
The solid black and solid magenta lines refer to the nominal tabulated results from the \textsc{Srim} database for electronic and ionic stopping powers respectively \cite{Ziegler_2010}. 
Open circles and open squares show data of Cantero \emph{et al.} \cite{Cantero_2009} and Markin \emph{et al.} \cite{Markin_2009} respectively. 
Dash (green) line and dash-dot (brown) line corresponds to a linear RPA calculation for a free-electron gas with $r_\text{s} = 2.67 ~a_0$ ($1$e per $\mathrm{Cu}$ atom) and $r_\text{s} = 1.12 ~a_0$ ($11$e per $\mathrm{Cu}$ atom). 
The thin solid black line, obtained by a linear fit to our simulated results, is the polynomial curve $\propto v^q$  to fit the crossover region (with a power of $q = 1.48 \pm 0.02$).
}}
\end{center}
\end{figure}

A logarithmic version of the findings of Fig. \ref{fig:stopping_power} is depicted in Fig.~\ref{fig:log_stopping_power}, where we have observed that the resulting curve is not as particularly simple.
In order to interpret the results we also calculated the linear response stopping $S_\text{L}(n, v)$ \cite{Lindhard_1964_book} based on a simple Lindhard RPA dielectric function $\varepsilon_\text{RPA}$ for different effective densities $n$ of the homogeneous electron gas~\cite{Giuliani_2005}
\begin{equation}
    S_\text{L}(n, v) = \frac{2 e^2}{\pi v^2} \int_0^\infty \frac{\mathrm{d}k}{k} \int_0^{k v} \omega\mathrm{d}\omega \Im\left(\frac{1}{\varepsilon_\text{RPA}(n, k, \omega)}\right)
\end{equation}
(which assumes a proton effective charge of $Z_1 = 1$).
As shown in Fig.~\ref{fig:log_stopping_power}, for $v < 0.07~\mathrm{a.u.}$ the response of the effective electron gas with one electron per $\mathrm{Cu}$ mimics the TDDFT results.
While more sophisticated dielectric models can be used \cite{Morawetz_1996}, we use the minimal model that can explain the simulation in the different regimes.

The resulting curves in Fig.~\ref{fig:log_stopping_power} shows that for $v > 0.3~\mathrm{a.u.}$ at least the $11$ electrons per atom (full valence) participates in the stopping electron gas within linear response.
We observe a $S_\text{e}$ kink around $v\sim 0.07~\mathrm{a.u.}$ due to a mixture of $\mathrm{d}$-band in the electronic density of states.
Similarly, according to this analysis of our new results, for $v \leq 0.07~\mathrm{a.u.}$, are primarily due to $\mathrm{s}$-band electrons within linear response. 
In the simulation we directly show a crossover region between the two linear regimes, and we find that the friction is in direct relation to the velocity with a power law with exponent $1.48 \pm 0.02$.
(In linear scale kinks are represented by changes of curvature, here logarithmic scale is more appropriate to discuss the physics.)

The kink we found at $v = 0.07~\mathrm{a.u.}$ can be explained by conservation laws in the effective homogeneous electron gas and general properties of electronic density of states in crystalline $\mathrm{Cu}$.
The minimum energy loss with maximum momentum transfer from an electron to an ion moving with velocity $v$ are respectively $2\hbar k_\text{F}v$ and $2\hbar k_\text{F}$ (plus small corrections of order $m_\text{e}/m_\text{p}$). 
Due to Pauli exclusion, only electrons in the energy range $E_\text{F} \pm 2\hbar k_\text{F} v$ can participate in the stopping process. 
Taking into account that DFT band structure predicts that the $\mathrm{d}$-band edge is $\Delta_\text{DFT} = 1.6~\mathrm{eV}$ below the Fermi energy (see for example, Fig.~3(a) in Ref.~\cite{Lin_2008}), 
that electron (band) effective mass close to $1$ and that $k_\text{F} = 0.72/a_0$ for the effective homogeneous electron gas of $\mathrm{Cu}$ $\mathrm{s}$-electrons \cite{Ashcroft_2003}, we can derive an approximate value of $v_\text{kink}$ caused by the participation of $\mathrm{d}$-electrons.
Based in this DFT ground state density of states plus conservation laws, we obtain an estimate of $v_\text{kink} = \Delta/(2\hbar k_\text{F}) = 0.41~\mathrm{a.u.}$ in near agreement with our TDDFT prediction.
In reality, the $\mathrm{d}$-band is about $\Delta_\text{exp} = 2~\mathrm{eV}$ below the Fermi energy as indicated by ARPES ~\cite{Knapp_1979}, which means that both the DFT-based estimate and the full TDDFT result should be giving an underestimation of 25\% of the kink location.
The second (negative) kink at $v = 0.3~\mathrm{a.u.}$ is more difficult to explain precisely  as the qualitative description in terms of $k_\text{F}$ (as in the homogeneous electron gas) becomes more ambiguous, but it is related to the point at which the whole conduction band ($11$ `$\mathrm{s} + \mathrm{d}$' electrons) starts participating in the process.

At low velocity, our results show good agreement with the experiments of Markin \emph{et al.} \cite{Markin_2009} but in relative disagreement with the measurements of Cantero \emph{et al.} \cite{Cantero_2009} and the \textsc{Srim} model. 
The difference between experiments could be due to a simple experimental scaling issue related to the difference between measuring relative and absolute stopping power at low velocities \cite{Markin_2009}.

Below $0.07~\mathrm{a.u.}$, the lack of experimental points precludes a direct comparison, but we find linear behavior at least down to $0.02~\mathrm{a.u.}$. 
Below $0.02~\mathrm{a.u.}$ the direct real time integration becomes less efficient and the accuracy is compromised by the quality of the numerical time integrator and the number of steps necessary to complete a calculation~\cite{Schleife_2012}.
Probing this regime experimentally would be rather difficult, especially to disentangle it from nuclear stopping effects, but if this is confirmed it would be an unexpected new regime.
In any case, the combination of experimental and theoretical results shows that the limit $v\to 0$ is intricate for metals as it is for insulators \cite{Artacho_2007, Lim_2016} where analogous band and gap threshold effects have been found.

Finally, we point out that the investigation of the low velocity limit of stopping power is important for the understanding of the non-adiabatic coupling between ions and electrons \cite{Caro_2015} and also for modeling dissipative molecular dynamics \cite{Caro_1989,Duffy_2006} where electrons act both as a thermal bath and a friction media. 
In simulations of radiation events the final state is precisely controlled by dissipation in the late stages when ions move slowly but still non-adibatically \cite{Zarkadoula_2014}.

\begin{figure}[h!]
\begin{center}
\includegraphics[width=0.531\columnwidth]{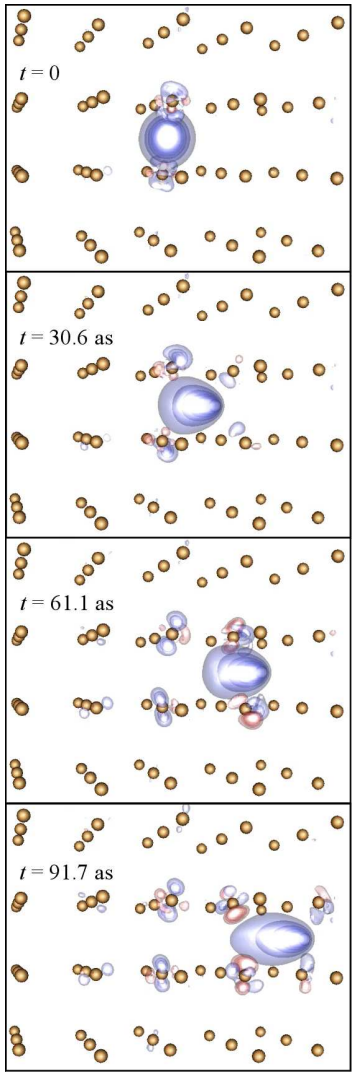} %%[width=0.5599999999999999\columnwidth]{figures/snapshot_copper/wake}
\caption{{\label{fig:snapshot_copper} (color online). 
Snapshots of density change produced by a $\mathrm{H^+}$ moving in $\mathrm{Cu}$ atoms with $v = 1.8 ~\mathrm{a.u.}$ (kinetic energy of $81 ~\mathrm{keV}$) along the $\langle 100\rangle$ channeling trajectory. 
The brown balls represents the $\mathrm{Cu}$ atoms and the single gray ball with light-blue iso-density contours represents host electron density affected by the presence of $\mathrm{H^+}$.
From initial condition ($t=0$) to a representative steady condition after $t = 91.65~\mathrm{attosecond}$ (see Fig. \ref{fig:energy_distance}).
Visualization produced with \textsc{VisIt}~\cite{Childs_2012}%
}}
\end{center}
\end{figure}

A snapshot of the redistribution of the charge density as a function of time for the simulation is shown in Fig.~\ref{fig:snapshot_copper}. At $t = 0~\text{attoseconds}$ (see Fig.~\ref{fig:snapshot_copper}a) the $\mathrm{H^+}$ interacts with the neighboring $\mathrm{Cu}$ atoms charge density where some electron density is acquired. 
Elongated wake tails are characteristic of projectiles traveling at or above the Fermi velocity~\cite{Tehada_1980}, in turn this affects the forces exerted on the neighboring atoms.
%. 

\begin{figure}[h!]
\begin{center}
\includegraphics[width=1\columnwidth]{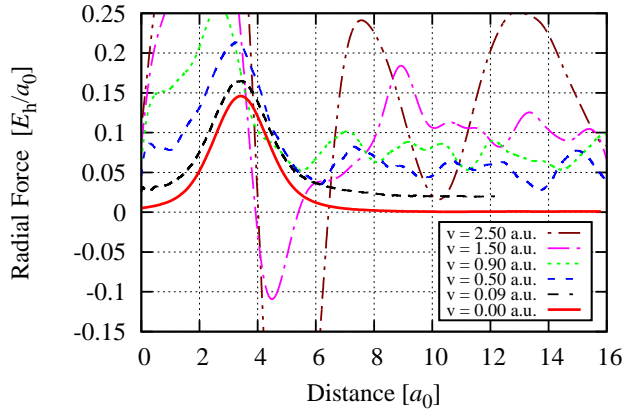}
\caption{{\label{fig:force_on_neighbor} (color online). Radial force exerted on a $\mathrm{Cu}$ atom as a function of parallel distance to projectile at different projectile velocities $v$. 
The force is \emph{negative} radial, which means that adiabatically (red curve) the proton attracts the neighbour $\mathrm{Cu}$ atom, but as the electron wake develops at higher velocities  ($v = 1.5$ and $2.5~\mathrm{a.u.}$ the force becomes repulsive after passing. 
For visualization purposes the non-adiabatic curves have been shifted vertically upwards.%%%
}}
\end{center}
\end{figure}

Fig. \ref{fig:force_on_neighbor} shows the radial force exerted on a neighboring $\mathrm{Cu}$ atom closest to $\mathrm{H^+}$ trajectory as a function of parallel distance to the projectile at different projectile velocities along the $\langle 100\rangle$ channeling trajectory. The forces on the nuclei are evaluated using the time-dependent electron density, $n(\mathbf{r}, t)$. 
%The force is evaluated by the Hellmann-Feynman theorem \cite{Feynman_1939,1941,Hellmann_2015}.

The adiabatic force is recovered for $v \to 0$ with no oscillations as expected. 
The maximum value for the force is obtained at the closest distance between the $\mathrm{H^+}$ and a neighbor $\mathrm{Cu}$ atom.
As the proton moves further from the $\mathrm{Cu}$ atom, the force decreases and eventually reduces to zero. 
As the velocity increases the position of the maximum value of the force first shifts and later results in persistent oscillations.
The existence of plasma oscillations is detected in our simulations by persistent charge motion above a certain threshold velocity of $v \simeq 1.0~\mathrm{a.u.}$. 
These plasma oscillations affect the components forces over individual $\mathrm{Cu}$ atoms near the trajectory of the passing hydrogen (Fig.~\ref{fig:force_on_neighbor}). These forces persist (and oscillate) even after the proton has passed. 
%These oscillations remain persistent with the increase of projectile velocity.% of the proton increases.%

\section{Conclusion}

In this paper we have reported the $S_\text{e}$ of protons in copper in a very wide range of velocities.
TDDFT-based electron dynamics is capable of capturing most of the physics in the different ranges, starting from non-linear screening effects, electron-hole excitations and production of plasmons. 
We disentangled channeling and off-channeling effects and observe a collapse of the two curves at low velocities; and identified four regimes
i) the linear $\mathrm{s}$-only ($0.02-0.07~\mathrm{a.u.}$), 
ii) linear $\mathrm{s}+\mathrm{d}$ ($0.3-1~\mathrm{a.u.}$), 
iii) crossover with $\sim 1.5$-power law ($0.07-0.3~\mathrm{a.u.}$) and iv) plasma-like ($v > 1~\mathrm{a.u.}$). 
%and v) what is possibly a non-linear screening regime at $v < 0.02~\mathrm{a.u.}$. 
This further illustrates that the electronic stopping in general does not possess a simple behavior in the limit $v\to 0$, and that band and bound effects dominate this behavior.

\section{Acknowledgements}
We would like to thank Professor C.~A. Weatherford, Dr. E.~R. Schwegler and Dr.~K.~J.~Reed for their keen interests, constant encouragements and useful suggestions throughout the progress of this work. This work was performed under the auspices of the U.S. Department of Energy by Lawrence Livermore National Laboratory under Contract DE-AC52-07NA27344.  Computing support for this work came from the Lawrence Livermore National Laboratory Institutional Computing Grand Challenge program. This is a collaborative project between the Department of Physics, Florida A\&M University and Lawrence Livermore National Laboratory. E. E. Quashie and B. C. Saha thankfully acknowledge the support by National Nuclear Security Administration (NNSA), Award Number(s) DE-NA0002630.% grant.%
%Department of Energy, National Nuclear Security Administration, Award Number(s) DE-NA0002630

\bibliographystyle{apsrev4-1}
%\bibliography{bibliography/converted_to_latex.bib%

\begin{thebibliography}{63}%
\makeatletter
\providecommand \@ifxundefined [1]{%
 \@ifx{#1\undefined}
}%
\providecommand \@ifnum [1]{%
 \ifnum #1\expandafter \@firstoftwo
 \else \expandafter \@secondoftwo
 \fi
}%
\providecommand \@ifx [1]{%
 \ifx #1\expandafter \@firstoftwo
 \else \expandafter \@secondoftwo
 \fi
}%
\providecommand \natexlab [1]{#1}%
\providecommand \enquote  [1]{``#1''}%
\providecommand \bibnamefont  [1]{#1}%
\providecommand \bibfnamefont [1]{#1}%
\providecommand \citenamefont [1]{#1}%
\providecommand \href@noop [0]{\@secondoftwo}%
\providecommand \href [0]{\begingroup \@sanitize@url \@href}%
\providecommand \@href[1]{\@@startlink{#1}\@@href}%
\providecommand \@@href[1]{\endgroup#1\@@endlink}%
\providecommand \@sanitize@url [0]{\catcode `\\12\catcode `\$12\catcode
  `\&12\catcode `\#12\catcode `\^12\catcode `\_12\catcode `\%12\relax}%
\providecommand \@@startlink[1]{}%
\providecommand \@@endlink[0]{}%
\providecommand \url  [0]{\begingroup\@sanitize@url \@url }%
\providecommand \@url [1]{\endgroup\@href {#1}{\urlprefix }}%
\providecommand \urlprefix  [0]{URL }%
\providecommand \Eprint [0]{\href }%
\providecommand \doibase [0]{http://dx.doi.org/}%
\providecommand \selectlanguage [0]{\@gobble}%
\providecommand \bibinfo  [0]{\@secondoftwo}%
\providecommand \bibfield  [0]{\@secondoftwo}%
\providecommand \translation [1]{[#1]}%
\providecommand \BibitemOpen [0]{}%
\providecommand \bibitemStop [0]{}%
\providecommand \bibitemNoStop [0]{.\EOS\space}%
\providecommand \EOS [0]{\spacefactor3000\relax}%
\providecommand \BibitemShut  [1]{\csname bibitem#1\endcsname}%
\let\auto@bib@innerbib\@empty
%</preamble>
\bibitem [{\citenamefont {Komarov}\ \emph {et~al.}(2013)\citenamefont
  {Komarov}, \citenamefont {Komarov}, \citenamefont {Pil'ko},\ and\
  \citenamefont {Pil'ko}}]{Komarov_2013}%
  \BibitemOpen
  \bibfield  {author} {\bibinfo {author} {\bibfnamefont {F.~F.}\ \bibnamefont
  {Komarov}}, \bibinfo {author} {\bibfnamefont {A.~F.}\ \bibnamefont
  {Komarov}}, \bibinfo {author} {\bibfnamefont {V.~V.}\ \bibnamefont {Pil'ko}},
  \ and\ \bibinfo {author} {\bibfnamefont {V.~V.}\ \bibnamefont {Pil'ko}},\
  }\href {\doibase 10.1007/s10891-013-0976-y} {\bibfield  {journal} {\bibinfo
  {journal} {J Eng Phys Thermophy}\ }\textbf {\bibinfo {volume} {86}},\
  \bibinfo {pages} {1481} (\bibinfo {year} {2013})}\BibitemShut {NoStop}%
\bibitem [{\citenamefont {Patel}\ \emph {et~al.}(2003)\citenamefont {Patel},
  \citenamefont {Mackinnon}, \citenamefont {Key}, \citenamefont {Cowan},
  \citenamefont {Foord}, \citenamefont {Allen}, \citenamefont {Price},
  \citenamefont {Ruhl}, \citenamefont {Springer},\ and\ \citenamefont
  {Stephens}}]{Patel_2003}%
  \BibitemOpen
  \bibfield  {author} {\bibinfo {author} {\bibfnamefont {P.}~\bibnamefont
  {Patel}}, \bibinfo {author} {\bibfnamefont {A.}~\bibnamefont {Mackinnon}},
  \bibinfo {author} {\bibfnamefont {M.}~\bibnamefont {Key}}, \bibinfo {author}
  {\bibfnamefont {T.}~\bibnamefont {Cowan}}, \bibinfo {author} {\bibfnamefont
  {M.}~\bibnamefont {Foord}}, \bibinfo {author} {\bibfnamefont
  {M.}~\bibnamefont {Allen}}, \bibinfo {author} {\bibfnamefont
  {D.}~\bibnamefont {Price}}, \bibinfo {author} {\bibfnamefont
  {H.}~\bibnamefont {Ruhl}}, \bibinfo {author} {\bibfnamefont {P.}~\bibnamefont
  {Springer}}, \ and\ \bibinfo {author} {\bibfnamefont {R.}~\bibnamefont
  {Stephens}},\ }\href {\doibase 10.1103/physrevlett.91.125004} {\bibfield
  {journal} {\bibinfo  {journal} {Phys. Rev. Lett.}\ }\textbf {\bibinfo
  {volume} {91}} (\bibinfo {year} {2003}),\
  10.1103/physrevlett.91.125004}\BibitemShut {NoStop}%
\bibitem [{\citenamefont {Caporaso}\ \emph {et~al.}(2009)\citenamefont
  {Caporaso}, \citenamefont {Chen},\ and\ \citenamefont
  {Sampayan}}]{Caporaso_2009}%
  \BibitemOpen
  \bibfield  {author} {\bibinfo {author} {\bibfnamefont {G.~J.}\ \bibnamefont
  {Caporaso}}, \bibinfo {author} {\bibfnamefont {Y.-J.}\ \bibnamefont {Chen}},
  \ and\ \bibinfo {author} {\bibfnamefont {S.~E.}\ \bibnamefont {Sampayan}},\
  }\href {\doibase 10.1142/s1793626809000235} {\bibfield  {journal} {\bibinfo
  {journal} {Reviews of Accelerator Science and Technology}\ }\textbf {\bibinfo
  {volume} {02}},\ \bibinfo {pages} {253} (\bibinfo {year} {2009})}\BibitemShut
  {NoStop}%
\bibitem [{\citenamefont {Odette}\ and\ \citenamefont
  {Wirth}(2005)}]{Odette_2005}%
  \BibitemOpen
  \bibfield  {author} {\bibinfo {author} {\bibfnamefont {G.~R.}\ \bibnamefont
  {Odette}}\ and\ \bibinfo {author} {\bibfnamefont {B.~D.}\ \bibnamefont
  {Wirth}},\ }in\ \href {\doibase 10.1007/1-4020-3286-2_50} {\emph {\bibinfo
  {booktitle} {Handbook of Materials Modeling}}}\ (\bibinfo  {publisher}
  {Springer Netherlands},\ \bibinfo {year} {2005})\ pp.\ \bibinfo {pages}
  {999--1037}\BibitemShut {NoStop}%
\bibitem [{\citenamefont {Ferrell}\ and\ \citenamefont
  {Ritchie}(1977)}]{Ferrell_1977}%
  \BibitemOpen
  \bibfield  {author} {\bibinfo {author} {\bibfnamefont {T.~L.}\ \bibnamefont
  {Ferrell}}\ and\ \bibinfo {author} {\bibfnamefont {R.~H.}\ \bibnamefont
  {Ritchie}},\ }\href {\doibase 10.1103/physrevb.16.115} {\bibfield  {journal}
  {\bibinfo  {journal} {Physical Review B}\ }\textbf {\bibinfo {volume} {16}},\
  \bibinfo {pages} {115} (\bibinfo {year} {1977})}\BibitemShut {NoStop}%
\bibitem [{\citenamefont {Chenakin}\ \emph {et~al.}(2006)\citenamefont
  {Chenakin}, \citenamefont {Markin}, \citenamefont {Steinbauer}, \citenamefont
  {Draxler},\ and\ \citenamefont {Bauer}}]{Chenakin_2006}%
  \BibitemOpen
  \bibfield  {author} {\bibinfo {author} {\bibfnamefont {S.}~\bibnamefont
  {Chenakin}}, \bibinfo {author} {\bibfnamefont {S.}~\bibnamefont {Markin}},
  \bibinfo {author} {\bibfnamefont {E.}~\bibnamefont {Steinbauer}}, \bibinfo
  {author} {\bibfnamefont {M.}~\bibnamefont {Draxler}}, \ and\ \bibinfo
  {author} {\bibfnamefont {P.}~\bibnamefont {Bauer}},\ }\href {\doibase
  10.1016/j.nimb.2006.03.023} {\bibfield  {journal} {\bibinfo  {journal}
  {Nuclear Instruments and Methods in Physics Research Section B: Beam
  Interactions with Materials and Atoms}\ }\textbf {\bibinfo {volume} {249}},\
  \bibinfo {pages} {58} (\bibinfo {year} {2006})}\BibitemShut {NoStop}%
\bibitem [{\citenamefont {Figueroa}\ \emph {et~al.}(2007)\citenamefont
  {Figueroa}, \citenamefont {Cantero}, \citenamefont {Eckardt}, \citenamefont
  {Lantschner}, \citenamefont {Vald{\'{e}}s},\ and\ \citenamefont
  {Arista}}]{Figueroa_2007}%
  \BibitemOpen
  \bibfield  {author} {\bibinfo {author} {\bibfnamefont {E.~A.}\ \bibnamefont
  {Figueroa}}, \bibinfo {author} {\bibfnamefont {E.~D.}\ \bibnamefont
  {Cantero}}, \bibinfo {author} {\bibfnamefont {J.~C.}\ \bibnamefont
  {Eckardt}}, \bibinfo {author} {\bibfnamefont {G.~H.}\ \bibnamefont
  {Lantschner}}, \bibinfo {author} {\bibfnamefont {J.~E.}\ \bibnamefont
  {Vald{\'{e}}s}}, \ and\ \bibinfo {author} {\bibfnamefont {N.~R.}\
  \bibnamefont {Arista}},\ }\href {\doibase 10.1103/physreva.75.019905}
  {\bibfield  {journal} {\bibinfo  {journal} {Phys. Rev. A}\ }\textbf {\bibinfo
  {volume} {75}} (\bibinfo {year} {2007}),\
  10.1103/physreva.75.019905}\BibitemShut {NoStop}%
\bibitem [{\citenamefont {Markin}\ \emph {et~al.}(2008)\citenamefont {Markin},
  \citenamefont {Primetzhofer}, \citenamefont {Prusa}, \citenamefont
  {Brunmayr}, \citenamefont {Kowarik}, \citenamefont {Aumayr},\ and\
  \citenamefont {Bauer}}]{Markin_2008}%
  \BibitemOpen
  \bibfield  {author} {\bibinfo {author} {\bibfnamefont {S.~N.}\ \bibnamefont
  {Markin}}, \bibinfo {author} {\bibfnamefont {D.}~\bibnamefont
  {Primetzhofer}}, \bibinfo {author} {\bibfnamefont {S.}~\bibnamefont {Prusa}},
  \bibinfo {author} {\bibfnamefont {M.}~\bibnamefont {Brunmayr}}, \bibinfo
  {author} {\bibfnamefont {G.}~\bibnamefont {Kowarik}}, \bibinfo {author}
  {\bibfnamefont {F.}~\bibnamefont {Aumayr}}, \ and\ \bibinfo {author}
  {\bibfnamefont {P.}~\bibnamefont {Bauer}},\ }\href {\doibase
  10.1103/physrevb.78.195122} {\bibfield  {journal} {\bibinfo  {journal} {Phys.
  Rev. B}\ }\textbf {\bibinfo {volume} {78}} (\bibinfo {year} {2008}),\
  10.1103/physrevb.78.195122}\BibitemShut {NoStop}%
\bibitem [{\citenamefont {Kaminsky}(1965)}]{Kaminsky_1965}%
  \BibitemOpen
  \bibfield  {author} {\bibinfo {author} {\bibfnamefont {M.}~\bibnamefont
  {Kaminsky}},\ }in\ \href {\doibase 10.1007/978-3-642-46025-8_10} {\emph
  {\bibinfo {booktitle} {Atomic and Ionic Impact Phenomena on Metal
  Surfaces}}}\ (\bibinfo  {publisher} {Springer Science $+$ Business Media},\
  \bibinfo {year} {1965})\ pp.\ \bibinfo {pages} {142--237}\BibitemShut
  {NoStop}%
\bibitem [{\citenamefont {Lehmann}\ and\ \citenamefont
  {Doran}(1978)}]{Lehmann_1978}%
  \BibitemOpen
  \bibfield  {author} {\bibinfo {author} {\bibfnamefont {C.}~\bibnamefont
  {Lehmann}}\ and\ \bibinfo {author} {\bibfnamefont {D.~G.}\ \bibnamefont
  {Doran}},\ }\href {\doibase 10.1063/1.2994975} {\bibfield  {journal}
  {\bibinfo  {journal} {Phys. Today}\ }\textbf {\bibinfo {volume} {31}},\
  \bibinfo {pages} {70} (\bibinfo {year} {1978})}\BibitemShut {NoStop}%
\bibitem [{\citenamefont {Sigmund}(2014)}]{Sigmund_2014}%
  \BibitemOpen
  \bibfield  {author} {\bibinfo {author} {\bibfnamefont {P.}~\bibnamefont
  {Sigmund}},\ }\href {\doibase 10.1007/978-3-319-05564-0} {\emph {\bibinfo
  {title} {{Particle Penetration and Radiation Effects Volume 2}}}}\ (\bibinfo
  {publisher} {Springer International Publishing},\ \bibinfo {year}
  {2014})\BibitemShut {NoStop}%
\bibitem [{\citenamefont {Nastasi}\ \emph {et~al.}(1996)\citenamefont
  {Nastasi}, \citenamefont {Mayer},\ and\ \citenamefont
  {Hirvonen}}]{Nastasi_1996}%
  \BibitemOpen
  \bibfield  {author} {\bibinfo {author} {\bibfnamefont {M.}~\bibnamefont
  {Nastasi}}, \bibinfo {author} {\bibfnamefont {J.}~\bibnamefont {Mayer}}, \
  and\ \bibinfo {author} {\bibfnamefont {J.~K.}\ \bibnamefont {Hirvonen}},\
  }in\ \href {\doibase 10.1017/cbo9780511565007.006} {\emph {\bibinfo
  {booktitle} {Ion{\textendash}solid Interactions}}}\ (\bibinfo  {publisher}
  {Cambridge University Press ({CUP})},\ \bibinfo {year} {1996})\ pp.\ \bibinfo
  {pages} {88--114}\BibitemShut {NoStop}%
\bibitem [{\citenamefont {Bethe}(1930{\natexlab{a}})}]{Bethe_1930_EN}%
  \BibitemOpen
  \bibfield  {author} {\bibinfo {author} {\bibfnamefont {H.}~\bibnamefont
  {Bethe}},\ }\href@noop {} {\bibfield  {journal} {\bibinfo  {journal} {Ann.
  Phys. (Leipzig) Lett.}\ }\textbf {\bibinfo {volume} {5}} (\bibinfo {year}
  {1930}{\natexlab{a}})}\BibitemShut {NoStop}%
\bibitem [{\citenamefont {Bloch}(1933)}]{Bloch_1933}%
  \BibitemOpen
  \bibfield  {author} {\bibinfo {author} {\bibfnamefont {F.}~\bibnamefont
  {Bloch}},\ }\href {\doibase 10.1002/andp.19334080303} {\bibfield  {journal}
  {\bibinfo  {journal} {Ann. Phys.}\ }\textbf {\bibinfo {volume} {408}},\
  \bibinfo {pages} {285} (\bibinfo {year} {1933})}\BibitemShut {NoStop}%
\bibitem [{\citenamefont {Fermi}\ and\ \citenamefont
  {Teller}(1947)}]{Fermi_1947}%
  \BibitemOpen
  \bibfield  {author} {\bibinfo {author} {\bibfnamefont {E.}~\bibnamefont
  {Fermi}}\ and\ \bibinfo {author} {\bibfnamefont {E.}~\bibnamefont {Teller}},\
  }\href {\doibase 10.1103/physrev.72.399} {\bibfield  {journal} {\bibinfo
  {journal} {Phys. Rev.}\ }\textbf {\bibinfo {volume} {72}},\ \bibinfo {pages}
  {399} (\bibinfo {year} {1947})}\BibitemShut {NoStop}%
\bibitem [{\citenamefont {Lindhard}\ and\ \citenamefont
  {Winther}(1964)}]{Lindhard_Winther}%
  \BibitemOpen
  \bibfield  {author} {\bibinfo {author} {\bibfnamefont {J.}~\bibnamefont
  {Lindhard}}\ and\ \bibinfo {author} {\bibfnamefont {A.}~\bibnamefont
  {Winther}},\ }\href@noop {} {\bibfield  {journal} {\bibinfo  {journal} {Kgl.
  Danske Videnskab. Selskab. Mat. Fys. Medd.}\ }\textbf {\bibinfo {volume}
  {34}},\ \bibinfo {pages} {1} (\bibinfo {year} {1964})}\BibitemShut {NoStop}%
\bibitem [{\citenamefont {Echenique}\ \emph {et~al.}(1989)\citenamefont
  {Echenique}, \citenamefont {Nagy},\ and\ \citenamefont
  {Arnau}}]{Echenique_1989}%
  \BibitemOpen
  \bibfield  {author} {\bibinfo {author} {\bibfnamefont {P.~M.}\ \bibnamefont
  {Echenique}}, \bibinfo {author} {\bibfnamefont {I.}~\bibnamefont {Nagy}}, \
  and\ \bibinfo {author} {\bibfnamefont {A.}~\bibnamefont {Arnau}},\ }\href
  {\doibase 10.1002/qua.560360854} {\bibfield  {journal} {\bibinfo  {journal}
  {Int. J. Quantum Chem.}\ }\textbf {\bibinfo {volume} {36}},\ \bibinfo {pages}
  {521} (\bibinfo {year} {1989})}\BibitemShut {NoStop}%
\bibitem [{\citenamefont {Bethe}(1930{\natexlab{b}})}]{Bethe_1930}%
  \BibitemOpen
  \bibfield  {author} {\bibinfo {author} {\bibfnamefont {H.}~\bibnamefont
  {Bethe}},\ }\href {\doibase 10.1002/andp.19303970303} {\bibfield  {journal}
  {\bibinfo  {journal} {Ann. Phys.}\ }\textbf {\bibinfo {volume} {397}},\
  \bibinfo {pages} {325} (\bibinfo {year} {1930}{\natexlab{b}})}\BibitemShut
  {NoStop}%
\bibitem [{\citenamefont {Pruneda}\ \emph {et~al.}(2007)\citenamefont
  {Pruneda}, \citenamefont {S{\'{a}}nchez-Portal}, \citenamefont {Arnau},
  \citenamefont {Juaristi},\ and\ \citenamefont {Artacho}}]{Pruneda_2007}%
  \BibitemOpen
  \bibfield  {author} {\bibinfo {author} {\bibfnamefont {J.~M.}\ \bibnamefont
  {Pruneda}}, \bibinfo {author} {\bibfnamefont {D.}~\bibnamefont
  {S{\'{a}}nchez-Portal}}, \bibinfo {author} {\bibfnamefont {A.}~\bibnamefont
  {Arnau}}, \bibinfo {author} {\bibfnamefont {J.~I.}\ \bibnamefont {Juaristi}},
  \ and\ \bibinfo {author} {\bibfnamefont {E.}~\bibnamefont {Artacho}},\ }\href
  {\doibase 10.1103/physrevlett.99.235501} {\bibfield  {journal} {\bibinfo
  {journal} {Phys. Rev. Lett.}\ }\textbf {\bibinfo {volume} {99}} (\bibinfo
  {year} {2007}),\ 10.1103/physrevlett.99.235501}\BibitemShut {NoStop}%
\bibitem [{\citenamefont {Schleife}\ \emph {et~al.}(2015)\citenamefont
  {Schleife}, \citenamefont {Kanai},\ and\ \citenamefont
  {Correa}}]{Schleife_2015}%
  \BibitemOpen
  \bibfield  {author} {\bibinfo {author} {\bibfnamefont {A.}~\bibnamefont
  {Schleife}}, \bibinfo {author} {\bibfnamefont {Y.}~\bibnamefont {Kanai}}, \
  and\ \bibinfo {author} {\bibfnamefont {A.~A.}\ \bibnamefont {Correa}},\
  }\href {\doibase 10.1103/physrevb.91.014306} {\bibfield  {journal} {\bibinfo
  {journal} {Phys. Rev. B}\ }\textbf {\bibinfo {volume} {91}} (\bibinfo {year}
  {2015}),\ 10.1103/physrevb.91.014306}\BibitemShut {NoStop}%
\bibitem [{\citenamefont {Ullah}\ \emph {et~al.}(2015)\citenamefont {Ullah},
  \citenamefont {Corsetti}, \citenamefont {S{\'{a}}nchez-Portal},\ and\
  \citenamefont {Artacho}}]{Ullah_2015}%
  \BibitemOpen
  \bibfield  {author} {\bibinfo {author} {\bibfnamefont {R.}~\bibnamefont
  {Ullah}}, \bibinfo {author} {\bibfnamefont {F.}~\bibnamefont {Corsetti}},
  \bibinfo {author} {\bibfnamefont {D.}~\bibnamefont {S{\'{a}}nchez-Portal}}, \
  and\ \bibinfo {author} {\bibfnamefont {E.}~\bibnamefont {Artacho}},\ }\href
  {\doibase 10.1103/physrevb.91.125203} {\bibfield  {journal} {\bibinfo
  {journal} {Phys. Rev. B}\ }\textbf {\bibinfo {volume} {91}} (\bibinfo {year}
  {2015}),\ 10.1103/physrevb.91.125203}\BibitemShut {NoStop}%
\bibitem [{\citenamefont {Race}\ \emph {et~al.}(2010)\citenamefont {Race},
  \citenamefont {Mason}, \citenamefont {Finnis}, \citenamefont {Foulkes},
  \citenamefont {Horsfield},\ and\ \citenamefont {Sutton}}]{Race_2010}%
  \BibitemOpen
  \bibfield  {author} {\bibinfo {author} {\bibfnamefont {C.~P.}\ \bibnamefont
  {Race}}, \bibinfo {author} {\bibfnamefont {D.~R.}\ \bibnamefont {Mason}},
  \bibinfo {author} {\bibfnamefont {M.~W.}\ \bibnamefont {Finnis}}, \bibinfo
  {author} {\bibfnamefont {W.~M.~C.}\ \bibnamefont {Foulkes}}, \bibinfo
  {author} {\bibfnamefont {A.~P.}\ \bibnamefont {Horsfield}}, \ and\ \bibinfo
  {author} {\bibfnamefont {A.~P.}\ \bibnamefont {Sutton}},\ }\href {\doibase
  10.1088/0034-4885/73/11/116501} {\bibfield  {journal} {\bibinfo  {journal}
  {Rep. Prog. Phys.}\ }\textbf {\bibinfo {volume} {73}},\ \bibinfo {pages}
  {116501} (\bibinfo {year} {2010})}\BibitemShut {NoStop}%
\bibitem [{\citenamefont {Quijada}\ \emph {et~al.}(2007)\citenamefont
  {Quijada}, \citenamefont {Borisov}, \citenamefont {Nagy}, \citenamefont
  {Mui{\~{n}}o},\ and\ \citenamefont {Echenique}}]{Quijada_2007}%
  \BibitemOpen
  \bibfield  {author} {\bibinfo {author} {\bibfnamefont {M.}~\bibnamefont
  {Quijada}}, \bibinfo {author} {\bibfnamefont {A.~G.}\ \bibnamefont
  {Borisov}}, \bibinfo {author} {\bibfnamefont {I.}~\bibnamefont {Nagy}},
  \bibinfo {author} {\bibfnamefont {R.~D.}\ \bibnamefont {Mui{\~{n}}o}}, \ and\
  \bibinfo {author} {\bibfnamefont {P.~M.}\ \bibnamefont {Echenique}},\ }\href
  {\doibase 10.1103/physreva.75.042902} {\bibfield  {journal} {\bibinfo
  {journal} {Phys. Rev. A}\ }\textbf {\bibinfo {volume} {75}} (\bibinfo {year}
  {2007}),\ 10.1103/physreva.75.042902}\BibitemShut {NoStop}%
\bibitem [{\citenamefont {Uddin}\ \emph {et~al.}(2013)\citenamefont {Uddin},
  \citenamefont {Haque}, \citenamefont {Talukder}, \citenamefont {Basak},
  \citenamefont {Saha},\ and\ \citenamefont {Malik}}]{Alfaz_Uddin_2013}%
  \BibitemOpen
  \bibfield  {author} {\bibinfo {author} {\bibfnamefont {M.~A.}\ \bibnamefont
  {Uddin}}, \bibinfo {author} {\bibfnamefont {A.~F.}\ \bibnamefont {Haque}},
  \bibinfo {author} {\bibfnamefont {T.~I.}\ \bibnamefont {Talukder}}, \bibinfo
  {author} {\bibfnamefont {A.~K.}\ \bibnamefont {Basak}}, \bibinfo {author}
  {\bibfnamefont {B.~C.}\ \bibnamefont {Saha}}, \ and\ \bibinfo {author}
  {\bibfnamefont {F.~B.}\ \bibnamefont {Malik}},\ }\href {\doibase
  10.1140/epjd/e2013-40164-8} {\bibfield  {journal} {\bibinfo  {journal} {The
  European Physical Journal D}\ }\textbf {\bibinfo {volume} {67}} (\bibinfo
  {year} {2013}),\ 10.1140/epjd/e2013-40164-8}\BibitemShut {NoStop}%
\bibitem [{\citenamefont {Haque}\ \emph {et~al.}(2015)\citenamefont {Haque},
  \citenamefont {Patoary}, \citenamefont {Uddin}, \citenamefont {Basak},\ and\
  \citenamefont {Saha}}]{Haque_2015}%
  \BibitemOpen
  \bibfield  {author} {\bibinfo {author} {\bibfnamefont {A.}~\bibnamefont
  {Haque}}, \bibinfo {author} {\bibfnamefont {M.~A.~R.}\ \bibnamefont
  {Patoary}}, \bibinfo {author} {\bibfnamefont {M.}~\bibnamefont {Uddin}},
  \bibinfo {author} {\bibfnamefont {A.}~\bibnamefont {Basak}}, \ and\ \bibinfo
  {author} {\bibfnamefont {B.}~\bibnamefont {Saha}},\ }\href {\doibase
  10.1080/00268976.2015.1074743} {\bibfield  {journal} {\bibinfo  {journal}
  {Molecular Physics}\ ,\ \bibinfo {pages} {1}} (\bibinfo {year}
  {2015})}\BibitemShut {NoStop}%
\bibitem [{\citenamefont {Lindhard}\ and\ \citenamefont
  {Scharff}(1953)}]{Lindhard_Scharff}%
  \BibitemOpen
  \bibfield  {author} {\bibinfo {author} {\bibfnamefont {J.}~\bibnamefont
  {Lindhard}}\ and\ \bibinfo {author} {\bibfnamefont {M.}~\bibnamefont
  {Scharff}},\ }\href@noop {} {\bibfield  {journal} {\bibinfo  {journal} {Kgl.
  Danske Videnskab. Selskab. Mat. Fys. Medd.}\ }\textbf {\bibinfo {volume}
  {27}},\ \bibinfo {pages} {15} (\bibinfo {year} {1953})}\BibitemShut {NoStop}%
\bibitem [{\citenamefont {Ziegler}\ \emph {et~al.}(2010)\citenamefont
  {Ziegler}, \citenamefont {Ziegler},\ and\ \citenamefont
  {Biersack}}]{Ziegler_2010}%
  \BibitemOpen
  \bibfield  {author} {\bibinfo {author} {\bibfnamefont {J.~F.}\ \bibnamefont
  {Ziegler}}, \bibinfo {author} {\bibfnamefont {M.}~\bibnamefont {Ziegler}}, \
  and\ \bibinfo {author} {\bibfnamefont {J.}~\bibnamefont {Biersack}},\ }\href
  {\doibase 10.1016/j.nimb.2010.02.091} {\bibfield  {journal} {\bibinfo
  {journal} {Nuclear Instruments and Methods in Physics Research Section B:
  Beam Interactions with Materials and Atoms}\ }\textbf {\bibinfo {volume}
  {268}},\ \bibinfo {pages} {1818} (\bibinfo {year} {2010})}\BibitemShut
  {NoStop}%
\bibitem [{\citenamefont {Correa}\ \emph {et~al.}(2012)\citenamefont {Correa},
  \citenamefont {Kohanoff}, \citenamefont {Artacho}, \citenamefont
  {Sanchez-Portal},\ and\ \citenamefont {Caro}}]{Correa_2012}%
  \BibitemOpen
  \bibfield  {author} {\bibinfo {author} {\bibfnamefont {A.~A.}\ \bibnamefont
  {Correa}}, \bibinfo {author} {\bibfnamefont {J.}~\bibnamefont {Kohanoff}},
  \bibinfo {author} {\bibfnamefont {E.}~\bibnamefont {Artacho}}, \bibinfo
  {author} {\bibfnamefont {D.}~\bibnamefont {Sanchez-Portal}}, \ and\ \bibinfo
  {author} {\bibfnamefont {A.}~\bibnamefont {Caro}},\ }\href {\doibase
  10.1103/physrevlett.109.069901} {\bibfield  {journal} {\bibinfo  {journal}
  {Phys. Rev. Lett.}\ }\textbf {\bibinfo {volume} {109}} (\bibinfo {year}
  {2012}),\ 10.1103/physrevlett.109.069901}\BibitemShut {NoStop}%
\bibitem [{\citenamefont {Cantero}\ \emph {et~al.}(2009)\citenamefont
  {Cantero}, \citenamefont {Lantschner}, \citenamefont {Eckardt},\ and\
  \citenamefont {Arista}}]{Cantero_2009}%
  \BibitemOpen
  \bibfield  {author} {\bibinfo {author} {\bibfnamefont {E.~D.}\ \bibnamefont
  {Cantero}}, \bibinfo {author} {\bibfnamefont {G.~H.}\ \bibnamefont
  {Lantschner}}, \bibinfo {author} {\bibfnamefont {J.~C.}\ \bibnamefont
  {Eckardt}}, \ and\ \bibinfo {author} {\bibfnamefont {N.~R.}\ \bibnamefont
  {Arista}},\ }\href {\doibase 10.1103/physreva.80.032904} {\bibfield
  {journal} {\bibinfo  {journal} {Phys. Rev. A}\ }\textbf {\bibinfo {volume}
  {80}} (\bibinfo {year} {2009}),\ 10.1103/physreva.80.032904}\BibitemShut
  {NoStop}%
\bibitem [{\citenamefont {Markin}\ \emph {et~al.}(2009)\citenamefont {Markin},
  \citenamefont {Primetzhofer}, \citenamefont {Spitz},\ and\ \citenamefont
  {Bauer}}]{Markin_2009}%
  \BibitemOpen
  \bibfield  {author} {\bibinfo {author} {\bibfnamefont {S.~N.}\ \bibnamefont
  {Markin}}, \bibinfo {author} {\bibfnamefont {D.}~\bibnamefont
  {Primetzhofer}}, \bibinfo {author} {\bibfnamefont {M.}~\bibnamefont {Spitz}},
  \ and\ \bibinfo {author} {\bibfnamefont {P.}~\bibnamefont {Bauer}},\ }\href
  {\doibase 10.1103/physrevb.80.205105} {\bibfield  {journal} {\bibinfo
  {journal} {Phys. Rev. B}\ }\textbf {\bibinfo {volume} {80}} (\bibinfo {year}
  {2009}),\ 10.1103/physrevb.80.205105}\BibitemShut {NoStop}%
\bibitem [{\citenamefont {Goebl}\ \emph {et~al.}(2013)\citenamefont {Goebl},
  \citenamefont {Roth},\ and\ \citenamefont {Bauer}}]{Goebl_2013}%
  \BibitemOpen
  \bibfield  {author} {\bibinfo {author} {\bibfnamefont {D.}~\bibnamefont
  {Goebl}}, \bibinfo {author} {\bibfnamefont {D.}~\bibnamefont {Roth}}, \ and\
  \bibinfo {author} {\bibfnamefont {P.}~\bibnamefont {Bauer}},\ }\href
  {\doibase 10.1103/physreva.87.062903} {\bibfield  {journal} {\bibinfo
  {journal} {Phys. Rev. A}\ }\textbf {\bibinfo {volume} {87}} (\bibinfo {year}
  {2013}),\ 10.1103/physreva.87.062903}\BibitemShut {NoStop}%
\bibitem [{\citenamefont {Schleife}\ \emph {et~al.}(2012)\citenamefont
  {Schleife}, \citenamefont {Draeger}, \citenamefont {Kanai},\ and\
  \citenamefont {Correa}}]{Schleife_2012}%
  \BibitemOpen
  \bibfield  {author} {\bibinfo {author} {\bibfnamefont {A.}~\bibnamefont
  {Schleife}}, \bibinfo {author} {\bibfnamefont {E.~W.}\ \bibnamefont
  {Draeger}}, \bibinfo {author} {\bibfnamefont {Y.}~\bibnamefont {Kanai}}, \
  and\ \bibinfo {author} {\bibfnamefont {A.~A.}\ \bibnamefont {Correa}},\
  }\href {\doibase 10.1063/1.4758792} {\bibfield  {journal} {\bibinfo
  {journal} {The Journal of Chemical Physics}\ }\textbf {\bibinfo {volume}
  {137}},\ \bibinfo {pages} {22A546} (\bibinfo {year} {2012})}\BibitemShut
  {NoStop}%
\bibitem [{\citenamefont {Schleife}\ \emph {et~al.}(2014)\citenamefont
  {Schleife}, \citenamefont {Draeger}, \citenamefont {Anisimov}, \citenamefont
  {Correa},\ and\ \citenamefont {Kanai}}]{Schleife_2014}%
  \BibitemOpen
  \bibfield  {author} {\bibinfo {author} {\bibfnamefont {A.}~\bibnamefont
  {Schleife}}, \bibinfo {author} {\bibfnamefont {E.~W.}\ \bibnamefont
  {Draeger}}, \bibinfo {author} {\bibfnamefont {V.~M.}\ \bibnamefont
  {Anisimov}}, \bibinfo {author} {\bibfnamefont {A.~A.}\ \bibnamefont
  {Correa}}, \ and\ \bibinfo {author} {\bibfnamefont {Y.}~\bibnamefont
  {Kanai}},\ }\href {\doibase 10.1109/mcse.2014.55} {\bibfield  {journal}
  {\bibinfo  {journal} {Comput. Sci. Eng.}\ }\textbf {\bibinfo {volume} {16}},\
  \bibinfo {pages} {54} (\bibinfo {year} {2014})}\BibitemShut {NoStop}%
\bibitem [{\citenamefont {Gross}\ \emph {et~al.}(1996)\citenamefont {Gross},
  \citenamefont {Dobson},\ and\ \citenamefont {Petersilka}}]{Gross_1996}%
  \BibitemOpen
  \bibfield  {author} {\bibinfo {author} {\bibfnamefont {E.~K.~U.}\
  \bibnamefont {Gross}}, \bibinfo {author} {\bibfnamefont {J.~F.}\ \bibnamefont
  {Dobson}}, \ and\ \bibinfo {author} {\bibfnamefont {M.}~\bibnamefont
  {Petersilka}},\ }in\ \href {\doibase 10.1007/bfb0016643} {\emph {\bibinfo
  {booktitle} {Density Functional Theory {II}}}}\ (\bibinfo  {publisher}
  {Springer Science $+$ Business Media},\ \bibinfo {year} {1996})\ pp.\
  \bibinfo {pages} {81--172}\BibitemShut {NoStop}%
\bibitem [{\citenamefont {Calvayrac}\ \emph {et~al.}(2000)\citenamefont
  {Calvayrac}, \citenamefont {Reinhard}, \citenamefont {Suraud},\ and\
  \citenamefont {Ullrich}}]{Calvayrac_2000}%
  \BibitemOpen
  \bibfield  {author} {\bibinfo {author} {\bibfnamefont {F.}~\bibnamefont
  {Calvayrac}}, \bibinfo {author} {\bibfnamefont {P.-G.}\ \bibnamefont
  {Reinhard}}, \bibinfo {author} {\bibfnamefont {E.}~\bibnamefont {Suraud}}, \
  and\ \bibinfo {author} {\bibfnamefont {C.}~\bibnamefont {Ullrich}},\ }\href
  {\doibase 10.1016/s0370-1573(00)00043-0} {\bibfield  {journal} {\bibinfo
  {journal} {Physics Reports}\ }\textbf {\bibinfo {volume} {337}},\ \bibinfo
  {pages} {493} (\bibinfo {year} {2000})}\BibitemShut {NoStop}%
\bibitem [{\citenamefont {Mason}\ \emph {et~al.}(2007)\citenamefont {Mason},
  \citenamefont {le~Page}, \citenamefont {Race}, \citenamefont {Foulkes},
  \citenamefont {Finnis},\ and\ \citenamefont {Sutton}}]{Mason_2007}%
  \BibitemOpen
  \bibfield  {author} {\bibinfo {author} {\bibfnamefont {D.~R.}\ \bibnamefont
  {Mason}}, \bibinfo {author} {\bibfnamefont {J.}~\bibnamefont {le~Page}},
  \bibinfo {author} {\bibfnamefont {C.~P.}\ \bibnamefont {Race}}, \bibinfo
  {author} {\bibfnamefont {W.~M.~C.}\ \bibnamefont {Foulkes}}, \bibinfo
  {author} {\bibfnamefont {M.~W.}\ \bibnamefont {Finnis}}, \ and\ \bibinfo
  {author} {\bibfnamefont {A.~P.}\ \bibnamefont {Sutton}},\ }\href {\doibase
  10.1088/0953-8984/19/43/436209} {\bibfield  {journal} {\bibinfo  {journal}
  {Journal of Physics: Condensed Matter}\ }\textbf {\bibinfo {volume} {19}},\
  \bibinfo {pages} {436209} (\bibinfo {year} {2007})}\BibitemShut {NoStop}%
\bibitem [{\citenamefont {Alonso}\ \emph {et~al.}(2008)\citenamefont {Alonso},
  \citenamefont {Andrade}, \citenamefont {Echenique}, \citenamefont {Falceto},
  \citenamefont {Prada-Gracia},\ and\ \citenamefont {Rubio}}]{Alonso_2008}%
  \BibitemOpen
  \bibfield  {author} {\bibinfo {author} {\bibfnamefont {J.~L.}\ \bibnamefont
  {Alonso}}, \bibinfo {author} {\bibfnamefont {X.}~\bibnamefont {Andrade}},
  \bibinfo {author} {\bibfnamefont {P.}~\bibnamefont {Echenique}}, \bibinfo
  {author} {\bibfnamefont {F.}~\bibnamefont {Falceto}}, \bibinfo {author}
  {\bibfnamefont {D.}~\bibnamefont {Prada-Gracia}}, \ and\ \bibinfo {author}
  {\bibfnamefont {A.}~\bibnamefont {Rubio}},\ }\href {\doibase
  10.1103/physrevlett.101.096403} {\bibfield  {journal} {\bibinfo  {journal}
  {Phys. Rev. Lett.}\ }\textbf {\bibinfo {volume} {101}} (\bibinfo {year}
  {2008}),\ 10.1103/physrevlett.101.096403}\BibitemShut {NoStop}%
\bibitem [{\citenamefont {Andrade}\ \emph {et~al.}(2009)\citenamefont
  {Andrade}, \citenamefont {Castro}, \citenamefont {Zueco}, \citenamefont
  {Alonso}, \citenamefont {Echenique}, \citenamefont {Falceto},\ and\
  \citenamefont {Rubio}}]{Andrade_2009}%
  \BibitemOpen
  \bibfield  {author} {\bibinfo {author} {\bibfnamefont {X.}~\bibnamefont
  {Andrade}}, \bibinfo {author} {\bibfnamefont {A.}~\bibnamefont {Castro}},
  \bibinfo {author} {\bibfnamefont {D.}~\bibnamefont {Zueco}}, \bibinfo
  {author} {\bibfnamefont {J.~L.}\ \bibnamefont {Alonso}}, \bibinfo {author}
  {\bibfnamefont {P.}~\bibnamefont {Echenique}}, \bibinfo {author}
  {\bibfnamefont {F.}~\bibnamefont {Falceto}}, \ and\ \bibinfo {author}
  {\bibfnamefont {A.}~\bibnamefont {Rubio}},\ }\href {\doibase
  10.1021/ct800518j} {\bibfield  {journal} {\bibinfo  {journal} {J. Chem.
  Theory Comput.}\ }\textbf {\bibinfo {volume} {5}},\ \bibinfo {pages} {728}
  (\bibinfo {year} {2009})}\BibitemShut {NoStop}%
\bibitem [{\citenamefont {Runge}\ and\ \citenamefont
  {Gross}(1984)}]{Runge_1984}%
  \BibitemOpen
  \bibfield  {author} {\bibinfo {author} {\bibfnamefont {E.}~\bibnamefont
  {Runge}}\ and\ \bibinfo {author} {\bibfnamefont {E.~K.~U.}\ \bibnamefont
  {Gross}},\ }\href {\doibase 10.1103/physrevlett.52.997} {\bibfield  {journal}
  {\bibinfo  {journal} {Phys. Rev. Lett.}\ }\textbf {\bibinfo {volume} {52}},\
  \bibinfo {pages} {997} (\bibinfo {year} {1984})}\BibitemShut {NoStop}%
\bibitem [{\citenamefont {Perdew}\ and\ \citenamefont
  {Wang}(1992)}]{Perdew_1992}%
  \BibitemOpen
  \bibfield  {author} {\bibinfo {author} {\bibfnamefont {J.~P.}\ \bibnamefont
  {Perdew}}\ and\ \bibinfo {author} {\bibfnamefont {Y.}~\bibnamefont {Wang}},\
  }\href {\doibase 10.1103/physrevb.45.13244} {\bibfield  {journal} {\bibinfo
  {journal} {Physical Review B}\ }\textbf {\bibinfo {volume} {45}},\ \bibinfo
  {pages} {13244} (\bibinfo {year} {1992})}\BibitemShut {NoStop}%
\bibitem [{\citenamefont {Perdew}\ \emph {et~al.}(1996)\citenamefont {Perdew},
  \citenamefont {Burke},\ and\ \citenamefont {Ernzerhof}}]{Perdew_1996}%
  \BibitemOpen
  \bibfield  {author} {\bibinfo {author} {\bibfnamefont {J.~P.}\ \bibnamefont
  {Perdew}}, \bibinfo {author} {\bibfnamefont {K.}~\bibnamefont {Burke}}, \
  and\ \bibinfo {author} {\bibfnamefont {M.}~\bibnamefont {Ernzerhof}},\ }\href
  {\doibase 10.1103/physrevlett.77.3865} {\bibfield  {journal} {\bibinfo
  {journal} {Phys. Rev. Lett.}\ }\textbf {\bibinfo {volume} {77}},\ \bibinfo
  {pages} {3865} (\bibinfo {year} {1996})}\BibitemShut {NoStop}%
\bibitem [{\citenamefont {Gygi}(2008)}]{Gygi_2008}%
  \BibitemOpen
  \bibfield  {author} {\bibinfo {author} {\bibfnamefont {F.}~\bibnamefont
  {Gygi}},\ }\href {\doibase 10.1147/rd.521.0137} {\bibfield  {journal}
  {\bibinfo  {journal} {{IBM} Journal of Research and Development}\ }\textbf
  {\bibinfo {volume} {52}},\ \bibinfo {pages} {137} (\bibinfo {year}
  {2008})}\BibitemShut {NoStop}%
\bibitem [{\citenamefont {Draeger}\ \emph {et~al.}(2016)\citenamefont
  {Draeger}, \citenamefont {Andrade}, \citenamefont {Gunnels}, \citenamefont
  {Bhatele}, \citenamefont {Schleife},\ and\ \citenamefont
  {Correa}}]{Draeger_2016}%
  \BibitemOpen
  \bibfield  {author} {\bibinfo {author} {\bibfnamefont {E.~W.}\ \bibnamefont
  {Draeger}}, \bibinfo {author} {\bibfnamefont {X.}~\bibnamefont {Andrade}},
  \bibinfo {author} {\bibfnamefont {J.~A.}\ \bibnamefont {Gunnels}}, \bibinfo
  {author} {\bibfnamefont {A.}~\bibnamefont {Bhatele}}, \bibinfo {author}
  {\bibfnamefont {A.}~\bibnamefont {Schleife}}, \ and\ \bibinfo {author}
  {\bibfnamefont {A.~A.}\ \bibnamefont {Correa}},\ }in\ \href {\doibase
  10.1109/ipdps.2016.46} {\emph {\bibinfo {booktitle} {2016 {IEEE}
  International Parallel and Distributed Processing Symposium ({IPDPS})}}}\
  (\bibinfo  {publisher} {Institute of Electrical {\&} amp $;$ Electronics
  Engineers ({IEEE})},\ \bibinfo {year} {2016})\BibitemShut {NoStop}%
\bibitem [{\citenamefont {Amisaki}(2000)}]{Amisaki_2000}%
  \BibitemOpen
  \bibfield  {author} {\bibinfo {author} {\bibfnamefont {T.}~\bibnamefont
  {Amisaki}},\ }\href {\doibase
  10.1002/1096-987x(200009)21:12<1075::aid-jcc4>3.0.co;2-l} {\bibfield
  {journal} {\bibinfo  {journal} {J. Comput. Chem.}\ }\textbf {\bibinfo
  {volume} {21}},\ \bibinfo {pages} {1075} (\bibinfo {year}
  {2000})}\BibitemShut {NoStop}%
\bibitem [{\citenamefont {Roy}(2007)}]{Roy_2007}%
  \BibitemOpen
  \bibfield  {author} {\bibinfo {author} {\bibfnamefont {S.~C.}\ \bibnamefont
  {Roy}},\ }in\ \href {\doibase 10.1533/9780857099426.77} {\emph {\bibinfo
  {booktitle} {Complex Numbers}}}\ (\bibinfo  {publisher} {Elsevier {BV}},\
  \bibinfo {year} {2007})\ pp.\ \bibinfo {pages} {77--114}\BibitemShut
  {NoStop}%
\bibitem [{\citenamefont {Pruneda}\ and\ \citenamefont
  {Artacho}(2005)}]{Pruneda_2005}%
  \BibitemOpen
  \bibfield  {author} {\bibinfo {author} {\bibfnamefont {J.~M.}\ \bibnamefont
  {Pruneda}}\ and\ \bibinfo {author} {\bibfnamefont {E.}~\bibnamefont
  {Artacho}},\ }\href {\doibase 10.1103/physrevb.71.094113} {\bibfield
  {journal} {\bibinfo  {journal} {Phys. Rev. B}\ }\textbf {\bibinfo {volume}
  {71}} (\bibinfo {year} {2005}),\ 10.1103/physrevb.71.094113}\BibitemShut
  {NoStop}%
\bibitem [{\citenamefont {Ziegler}()}]{Copper_Stop}%
  \BibitemOpen
  \bibfield  {author} {\bibinfo {author} {\bibfnamefont {J.~F.}\ \bibnamefont
  {Ziegler}},\ }\href {\doibase
  http://www.srim.org/SRIM/SRIMPICS/\\STOP01/STOP0129.gif} {\
  http://www.srim.org/SRIM/SRIMPICS/\\STOP01/STOP0129.gif}\BibitemShut
  {NoStop}%
\bibitem [{\citenamefont {Dorado}\ and\ \citenamefont
  {Flores}(1993)}]{Dorado_1993}%
  \BibitemOpen
  \bibfield  {author} {\bibinfo {author} {\bibfnamefont {J.~J.}\ \bibnamefont
  {Dorado}}\ and\ \bibinfo {author} {\bibfnamefont {F.}~\bibnamefont
  {Flores}},\ }\href {\doibase 10.1103/physreva.47.3062} {\bibfield  {journal}
  {\bibinfo  {journal} {Phys. Rev. A}\ }\textbf {\bibinfo {volume} {47}},\
  \bibinfo {pages} {3062} (\bibinfo {year} {1993})}\BibitemShut {NoStop}%
\bibitem [{\citenamefont {Nazarov}\ \emph {et~al.}(2007)\citenamefont
  {Nazarov}, \citenamefont {Pitarke}, \citenamefont {Takada}, \citenamefont
  {Vignale},\ and\ \citenamefont {Chang}}]{Nazarov_2007}%
  \BibitemOpen
  \bibfield  {author} {\bibinfo {author} {\bibfnamefont {V.~U.}\ \bibnamefont
  {Nazarov}}, \bibinfo {author} {\bibfnamefont {J.~M.}\ \bibnamefont
  {Pitarke}}, \bibinfo {author} {\bibfnamefont {Y.}~\bibnamefont {Takada}},
  \bibinfo {author} {\bibfnamefont {G.}~\bibnamefont {Vignale}}, \ and\
  \bibinfo {author} {\bibfnamefont {Y.-C.}\ \bibnamefont {Chang}},\ }\href
  {\doibase 10.1103/physrevb.76.205103} {\bibfield  {journal} {\bibinfo
  {journal} {Phys. Rev. B}\ }\textbf {\bibinfo {volume} {76}} (\bibinfo {year}
  {2007}),\ 10.1103/physrevb.76.205103}\BibitemShut {NoStop}%
\bibitem [{\citenamefont {Lindhard}\ \emph {et~al.}(1964)\citenamefont
  {Lindhard}, \citenamefont {Winther},\ and\ \citenamefont {videnskabernes
  selskab}}]{Lindhard_1964_book}%
  \BibitemOpen
  \bibfield  {author} {\bibinfo {author} {\bibfnamefont {J.}~\bibnamefont
  {Lindhard}}, \bibinfo {author} {\bibfnamefont {A.}~\bibnamefont {Winther}}, \
  and\ \bibinfo {author} {\bibfnamefont {K.~D.}\ \bibnamefont {videnskabernes
  selskab}},\ }\href {https://books.google.com/books?id=8kK\_tgAACAAJ} {\emph
  {\bibinfo {title} {{Stopping Power of Electron Gas and Equipartition
  Rule}}}},\ Mathematisk-fysiske meddelelser\ (\bibinfo  {publisher}
  {Munksgaard},\ \bibinfo {year} {1964})\BibitemShut {NoStop}%
\bibitem [{\citenamefont {Giuliani}\ and\ \citenamefont
  {Vignale}(2005)}]{Giuliani_2005}%
  \BibitemOpen
  \bibfield  {author} {\bibinfo {author} {\bibfnamefont {G.}~\bibnamefont
  {Giuliani}}\ and\ \bibinfo {author} {\bibfnamefont {G.}~\bibnamefont
  {Vignale}},\ }in\ \href {\doibase 10.1017/cbo9780511619915.006} {\emph
  {\bibinfo {booktitle} {Quantum Theory of the Electron Liquid}}}\ (\bibinfo
  {publisher} {Cambridge University Press ({CUP})},\ \bibinfo {year} {2005})\
  pp.\ \bibinfo {pages} {188--274}\BibitemShut {NoStop}%
\bibitem [{\citenamefont {Morawetz}\ and\ \citenamefont
  {Röpke}(1996)}]{Morawetz_1996}%
  \BibitemOpen
  \bibfield  {author} {\bibinfo {author} {\bibfnamefont {K.}~\bibnamefont
  {Morawetz}}\ and\ \bibinfo {author} {\bibfnamefont {G.}~\bibnamefont
  {Röpke}},\ }\href {\doibase 10.1103/physreve.54.4134} {\bibfield  {journal}
  {\bibinfo  {journal} {Physical Review E}\ }\textbf {\bibinfo {volume} {54}},\
  \bibinfo {pages} {4134} (\bibinfo {year} {1996})}\BibitemShut {NoStop}%
\bibitem [{\citenamefont {Lin}\ \emph {et~al.}(2008)\citenamefont {Lin},
  \citenamefont {Zhigilei},\ and\ \citenamefont {Celli}}]{Lin_2008}%
  \BibitemOpen
  \bibfield  {author} {\bibinfo {author} {\bibfnamefont {Z.}~\bibnamefont
  {Lin}}, \bibinfo {author} {\bibfnamefont {L.~V.}\ \bibnamefont {Zhigilei}}, \
  and\ \bibinfo {author} {\bibfnamefont {V.}~\bibnamefont {Celli}},\ }\href
  {\doibase 10.1103/physrevb.77.075133} {\bibfield  {journal} {\bibinfo
  {journal} {Phys. Rev. B}\ }\textbf {\bibinfo {volume} {77}} (\bibinfo {year}
  {2008}),\ 10.1103/physrevb.77.075133}\BibitemShut {NoStop}%
\bibitem [{\citenamefont {Ashcroft}\ and\ \citenamefont
  {Mermin}(1976)}]{Ashcroft_2003}%
  \BibitemOpen
  \bibfield  {author} {\bibinfo {author} {\bibfnamefont {N.}~\bibnamefont
  {Ashcroft}}\ and\ \bibinfo {author} {\bibfnamefont {N.}~\bibnamefont
  {Mermin}},\ }\href@noop {} {\emph {\bibinfo {title} {{Solid State
  Physics}}}}\ (\bibinfo  {publisher} {Saunders College},\ \bibinfo {address}
  {Philadelphia},\ \bibinfo {year} {1976})\BibitemShut {NoStop}%
\bibitem [{\citenamefont {Knapp}\ \emph {et~al.}(1979)\citenamefont {Knapp},
  \citenamefont {Himpsel},\ and\ \citenamefont {Eastman}}]{Knapp_1979}%
  \BibitemOpen
  \bibfield  {author} {\bibinfo {author} {\bibfnamefont {J.~A.}\ \bibnamefont
  {Knapp}}, \bibinfo {author} {\bibfnamefont {F.~J.}\ \bibnamefont {Himpsel}},
  \ and\ \bibinfo {author} {\bibfnamefont {D.~E.}\ \bibnamefont {Eastman}},\
  }\href {\doibase 10.1103/physrevb.19.4952} {\bibfield  {journal} {\bibinfo
  {journal} {Phys. Rev. B}\ }\textbf {\bibinfo {volume} {19}},\ \bibinfo
  {pages} {4952} (\bibinfo {year} {1979})}\BibitemShut {NoStop}%
\bibitem [{\citenamefont {Artacho}(2007)}]{Artacho_2007}%
  \BibitemOpen
  \bibfield  {author} {\bibinfo {author} {\bibfnamefont {E.}~\bibnamefont
  {Artacho}},\ }\href {\doibase 10.1088/0953-8984/19/27/275211} {\bibfield
  {journal} {\bibinfo  {journal} {Journal of Physics: Condensed Matter}\
  }\textbf {\bibinfo {volume} {19}},\ \bibinfo {pages} {275211} (\bibinfo
  {year} {2007})}\BibitemShut {NoStop}%
\bibitem [{\citenamefont {Lim}\ \emph {et~al.}(2016)\citenamefont {Lim},
  \citenamefont {Foulkes}, \citenamefont {Horsfield}, \citenamefont {Mason},
  \citenamefont {Schleife}, \citenamefont {Draeger},\ and\ \citenamefont
  {Correa}}]{Lim_2016}%
  \BibitemOpen
  \bibfield  {author} {\bibinfo {author} {\bibfnamefont {A.}~\bibnamefont
  {Lim}}, \bibinfo {author} {\bibfnamefont {W.~M.~C.}\ \bibnamefont {Foulkes}},
  \bibinfo {author} {\bibfnamefont {A.~P.}\ \bibnamefont {Horsfield}}, \bibinfo
  {author} {\bibfnamefont {D.~R.}\ \bibnamefont {Mason}}, \bibinfo {author}
  {\bibfnamefont {A.}~\bibnamefont {Schleife}}, \bibinfo {author}
  {\bibfnamefont {E.~W.}\ \bibnamefont {Draeger}}, \ and\ \bibinfo {author}
  {\bibfnamefont {A.~A.}\ \bibnamefont {Correa}},\ }\href {\doibase
  10.1103/physrevlett.116.043201} {\bibfield  {journal} {\bibinfo  {journal}
  {Phys. Rev. Lett.}\ }\textbf {\bibinfo {volume} {116}} (\bibinfo {year}
  {2016}),\ 10.1103/physrevlett.116.043201}\BibitemShut {NoStop}%
\bibitem [{\citenamefont {Caro}\ \emph {et~al.}(2015)\citenamefont {Caro},
  \citenamefont {Correa}, \citenamefont {Tamm}, \citenamefont {Samolyuk},\ and\
  \citenamefont {Stocks}}]{Caro_2015}%
  \BibitemOpen
  \bibfield  {author} {\bibinfo {author} {\bibfnamefont {A.}~\bibnamefont
  {Caro}}, \bibinfo {author} {\bibfnamefont {A.~A.}\ \bibnamefont {Correa}},
  \bibinfo {author} {\bibfnamefont {A.}~\bibnamefont {Tamm}}, \bibinfo {author}
  {\bibfnamefont {G.~D.}\ \bibnamefont {Samolyuk}}, \ and\ \bibinfo {author}
  {\bibfnamefont {G.~M.}\ \bibnamefont {Stocks}},\ }\href {\doibase
  10.1103/physrevb.92.144309} {\bibfield  {journal} {\bibinfo  {journal} {Phys.
  Rev. B}\ }\textbf {\bibinfo {volume} {92}} (\bibinfo {year} {2015}),\
  10.1103/physrevb.92.144309}\BibitemShut {NoStop}%
\bibitem [{\citenamefont {Caro}\ and\ \citenamefont
  {Victoria}(1989)}]{Caro_1989}%
  \BibitemOpen
  \bibfield  {author} {\bibinfo {author} {\bibfnamefont {A.}~\bibnamefont
  {Caro}}\ and\ \bibinfo {author} {\bibfnamefont {M.}~\bibnamefont
  {Victoria}},\ }\href {\doibase 10.1103/physreva.40.2287} {\bibfield
  {journal} {\bibinfo  {journal} {Phys. Rev. A}\ }\textbf {\bibinfo {volume}
  {40}},\ \bibinfo {pages} {2287} (\bibinfo {year} {1989})}\BibitemShut
  {NoStop}%
\bibitem [{\citenamefont {Duffy}\ and\ \citenamefont
  {Rutherford}(2006)}]{Duffy_2006}%
  \BibitemOpen
  \bibfield  {author} {\bibinfo {author} {\bibfnamefont {D.~M.}\ \bibnamefont
  {Duffy}}\ and\ \bibinfo {author} {\bibfnamefont {A.~M.}\ \bibnamefont
  {Rutherford}},\ }\href {\doibase 10.1088/0953-8984/19/1/016207} {\bibfield
  {journal} {\bibinfo  {journal} {Journal of Physics: Condensed Matter}\
  }\textbf {\bibinfo {volume} {19}},\ \bibinfo {pages} {016207} (\bibinfo
  {year} {2006})}\BibitemShut {NoStop}%
\bibitem [{\citenamefont {Zarkadoula}\ \emph {et~al.}(2014)\citenamefont
  {Zarkadoula}, \citenamefont {Daraszewicz}, \citenamefont {Duffy},
  \citenamefont {Seaton}, \citenamefont {Todorov}, \citenamefont {Nordlund},
  \citenamefont {Dove},\ and\ \citenamefont {Trachenko}}]{Zarkadoula_2014}%
  \BibitemOpen
  \bibfield  {author} {\bibinfo {author} {\bibfnamefont {E.}~\bibnamefont
  {Zarkadoula}}, \bibinfo {author} {\bibfnamefont {S.~L.}\ \bibnamefont
  {Daraszewicz}}, \bibinfo {author} {\bibfnamefont {D.~M.}\ \bibnamefont
  {Duffy}}, \bibinfo {author} {\bibfnamefont {M.~A.}\ \bibnamefont {Seaton}},
  \bibinfo {author} {\bibfnamefont {I.~T.}\ \bibnamefont {Todorov}}, \bibinfo
  {author} {\bibfnamefont {K.}~\bibnamefont {Nordlund}}, \bibinfo {author}
  {\bibfnamefont {M.~T.}\ \bibnamefont {Dove}}, \ and\ \bibinfo {author}
  {\bibfnamefont {K.}~\bibnamefont {Trachenko}},\ }\href {\doibase
  10.1088/0953-8984/26/8/085401} {\bibfield  {journal} {\bibinfo  {journal}
  {Journal of Physics: Condensed Matter}\ }\textbf {\bibinfo {volume} {26}},\
  \bibinfo {pages} {085401} (\bibinfo {year} {2014})}\BibitemShut {NoStop}%
\bibitem [{\citenamefont {Childs}\ \emph {et~al.}(2012)\citenamefont {Childs},
  \citenamefont {Brugger}, \citenamefont {Whitlock}, \citenamefont {Meredith},
  \citenamefont {Ahern}, \citenamefont {Pugmire}, \citenamefont {Biagas},
  \citenamefont {Miller}, \citenamefont {Harrison}, \citenamefont {Weber},
  \citenamefont {Krishnan}, \citenamefont {Fogal}, \citenamefont {Sanderson},
  \citenamefont {Garth}, \citenamefont {Bethel}, \citenamefont {Camp},
  \citenamefont {Rubel}, \citenamefont {Durant}, \citenamefont {Favre},\ and\
  \citenamefont {Navratil}}]{Childs_2012}%
  \BibitemOpen
  \bibfield  {author} {\bibinfo {author} {\bibfnamefont {H.}~\bibnamefont
  {Childs}}, \bibinfo {author} {\bibfnamefont {E.}~\bibnamefont {Brugger}},
  \bibinfo {author} {\bibfnamefont {B.}~\bibnamefont {Whitlock}}, \bibinfo
  {author} {\bibfnamefont {J.}~\bibnamefont {Meredith}}, \bibinfo {author}
  {\bibfnamefont {S.}~\bibnamefont {Ahern}}, \bibinfo {author} {\bibfnamefont
  {D.}~\bibnamefont {Pugmire}}, \bibinfo {author} {\bibfnamefont
  {K.}~\bibnamefont {Biagas}}, \bibinfo {author} {\bibfnamefont
  {M.}~\bibnamefont {Miller}}, \bibinfo {author} {\bibfnamefont
  {C.}~\bibnamefont {Harrison}}, \bibinfo {author} {\bibfnamefont
  {G.}~\bibnamefont {Weber}}, \bibinfo {author} {\bibfnamefont
  {H.}~\bibnamefont {Krishnan}}, \bibinfo {author} {\bibfnamefont
  {T.}~\bibnamefont {Fogal}}, \bibinfo {author} {\bibfnamefont
  {A.}~\bibnamefont {Sanderson}}, \bibinfo {author} {\bibfnamefont
  {C.}~\bibnamefont {Garth}}, \bibinfo {author} {\bibfnamefont
  {E.}~\bibnamefont {Bethel}}, \bibinfo {author} {\bibfnamefont
  {D.}~\bibnamefont {Camp}}, \bibinfo {author} {\bibfnamefont {O.}~\bibnamefont
  {Rubel}}, \bibinfo {author} {\bibfnamefont {M.}~\bibnamefont {Durant}},
  \bibinfo {author} {\bibfnamefont {J.}~\bibnamefont {Favre}}, \ and\ \bibinfo
  {author} {\bibfnamefont {P.}~\bibnamefont {Navratil}},\ }in\ \href {\doibase
  10.1201/b12985-21} {\emph {\bibinfo {booktitle} {High Performance
  Visualization}}}\ (\bibinfo  {publisher} {Informa {UK} Limited},\ \bibinfo
  {year} {2012})\BibitemShut {NoStop}%
\bibitem [{\citenamefont {Tehada}\ \emph {et~al.}(1980)\citenamefont {Tehada},
  \citenamefont {Echenique}, \citenamefont {Crawford},\ and\ \citenamefont
  {Ritchie}}]{Tehada_1980}%
  \BibitemOpen
  \bibfield  {author} {\bibinfo {author} {\bibfnamefont {J.}~\bibnamefont
  {Tehada}}, \bibinfo {author} {\bibfnamefont {P.}~\bibnamefont {Echenique}},
  \bibinfo {author} {\bibfnamefont {O.}~\bibnamefont {Crawford}}, \ and\
  \bibinfo {author} {\bibfnamefont {R.}~\bibnamefont {Ritchie}},\ }\href
  {\doibase 10.1016/0029-554x(80)91020-4} {\bibfield  {journal} {\bibinfo
  {journal} {Nuclear Instruments and Methods}\ }\textbf {\bibinfo {volume}
  {170}},\ \bibinfo {pages} {249} (\bibinfo {year} {1980})}\BibitemShut
  {NoStop}%
\end{thebibliography}
%}


\end{document}